\documentclass[preprint,superscriptaddress,floatfix,pra,amssymb,amsmath]{revtex4}

\usepackage{times,amsmath}
\usepackage{color}
\usepackage{graphicx}

\newcommand{\fn}{\zeta}

\newcommand{\Xin}{\delta X_{\mathrm{in}}}
\newcommand{\Yin}{\delta Y_{\mathrm{in}}}
\newcommand{\Xo}{\delta X_{\mathrm{out}}}
\newcommand{\Yo}{\delta Y_{\mathrm{out}}}
\newcommand{\XE}{X_{\mathrm{E}}}
\newcommand{\chio}{\chi_{\mathrm{opt}}}
\newcommand{\Gba}{\Gamma_{\mathrm{BA}}}
\newcommand{\Gth}{\Gamma_{\mathrm{th}}}
\newcommand{\Sth}{S_{q_\mathrm{th} q_\mathrm{th}}}
\newcommand{\rmi}{\mathrm{i}}

\newcommand{\qr}{q_\mathrm{r}}
\newcommand{\phio}{\phi_{\mathrm{opt}}}
\newcommand{\Coop}{\mathcal{C}}

\begin{document} 

\title{Quantum nondemolition measurement of optical field fluctuations by optomechanical interaction}

\author{A. Pontin \footnote{Present address: Department of Physics and Astronomy, University College London, WC1E 6BT, United Kingdom}}
\affiliation{Dipartimento di Fisica e Astronomia, Universit\`a di Firenze, Via Sansone 1, I-50019 Sesto Fiorentino (FI), Italy}
\affiliation{Istituto Nazionale di Fisica Nucleare (INFN), Sezione di Firenze, Via Sansone 1, I-50019 Sesto Fiorentino (FI), Italy}

\author{M. Bonaldi}
\affiliation{Institute of Materials for Electronics and Magnetism, Nanoscience-Trento-FBK Division,
 38123 Povo, Trento, Italy}
\affiliation{INFN, Trento Institute for Fundamental Physics and Application, I-38123 Povo, Trento, Italy }

\author{A. Borrielli}
\affiliation{Institute of Materials for Electronics and Magnetism, Nanoscience-Trento-FBK Division,
 38123 Povo, Trento, Italy}
\affiliation{INFN, Trento Institute for Fundamental Physics and Application, I-38123 Povo, Trento, Italy }

\author{L. Marconi}
\affiliation{CNR-INO, L.go Enrico Fermi 6, I-50125 Firenze, Italy}

\author{F. Marino}
\affiliation{Istituto Nazionale di Fisica Nucleare (INFN), Sezione di Firenze, Via Sansone 1, I-50019 Sesto Fiorentino (FI), Italy}
\affiliation{CNR-INO, L.go Enrico Fermi 6, I-50125 Firenze, Italy}

\author{G. Pandraud}
\affiliation{Delft University of Technology, Else Kooi Laboratory, 2628 Delft, The Netherlands}

\author{G. A. Prodi}
\affiliation{INFN, Trento Institute for Fundamental Physics and Application, I-38123 Povo, Trento, Italy }
\affiliation{Dipartimento di Fisica, Universit\`a di Trento, I-38123 Povo, Trento, Italy}

\author{P.M. Sarro}
\affiliation{Delft University of Technology, Else Kooi Laboratory, 2628 Delft, The Netherlands}

\author{E. Serra}
\affiliation{INFN, Trento Institute for Fundamental Physics and Application, I-38123 Povo, Trento, Italy }
\affiliation{Delft University of Technology, Else Kooi Laboratory, 2628 Delft, The Netherlands}

\author{F. Marin}
\email[Electronic mail: ]{marin@fi.infn.it}
\affiliation{Dipartimento di Fisica e Astronomia, Universit\`a di Firenze, Via Sansone 1, I-50019 Sesto Fiorentino (FI), Italy}
\affiliation{CNR-INO, L.go Enrico Fermi 6, I-50125 Firenze, Italy}
\affiliation{INFN, Sezione di Firenze, Via Sansone 1, I-50019 Sesto Fiorentino (FI), Italy}
\affiliation{European Laboratory for Non-Linear Spectroscopy (LENS), Via Carrara 1, I-50019 Sesto Fiorentino (FI), Italy}

\begin{abstract}
According to quantum mechanics, if we keep observing a continuous variable we generally disturb its evolution. For a class of observables, however, it is possible to implement a so-called quantum nondemolition measurement: by confining the perturbation to the conjugate variable, the observable is estimated with arbitrary accuracy, or prepared in a well-known state. For instance, when the light bounces on a movable mirror, its intensity is not perturbed (the effect is just seen on the phase of the radiation), but the radiation pressure allows to trace back its fluctuations by observing the mirror motion. In this work, we implement a cavity optomechanical experiment based on an oscillating micro-mirror, and we measure correlations between the output light intensity fluctuations and the mirror motion. We demonstrate that the uncertainty of the former is reduced below the shot noise level determined by the corpuscular nature of light.  
\end{abstract}

\maketitle

\section{Introduction}

Quantum mechanics generally prescribes that, as soon as we observe a system, we actually perturb it. 
As a paradigmatic example, in the Heisenberg's microscope a measurement of the position of a particle at the time $t$ perturbs its momentum, thus influencing the particle motion, and actually its position at following times. The consequence of the observation of the system (back-action) deteriorates the accuracy of a continuous measurement on the observable considered (the position). On the other hand, there are observables that are not affected by the disturbance caused by their measurement, the effect of which remains confined to their conjugate variable: their measurement can evade the back-action. For such observables it has been introduced the concept of Quantum Non-Demolition (QND) measurement \cite{ref1,ref2,ref3,ref4}. A QND measurement allows to keep observing a variable with arbitrary accuracy. Examples of QND variables are the quadratures of a mechanical oscillator and, similarly, the fluctuations on the quadratures of the electromagnetic field, defined from its bosonic operators, after separation of their average coherent amplitude ($a = \langle a \rangle + \delta a$), as $\delta X = \delta a+\delta a^{\dagger}$  (amplitude quadrature),  $\delta Y = -i(\delta a-\delta a^{\dagger})$ (phase quadrature)  and  $\delta X^{\phi} = \delta X \cos \phi + \delta Y \sin \phi$  (generic quadrature). 

The possibility to perform a QND measurement of a field quadrature (in particular, of the amplitude $\delta X$) by exploiting the radiation pressure exerted on a movable mirror was studied in a seminal work by Jacobs {\it et al.} in 1994 \cite{ref5}. 
When the light bounces on a mirror, its intensity is not perturbed: the displacement of the mirror changes the phase of the field, and the optomechanical interaction modifies $\delta Y$, but it leaves $\delta X$ unaffected. 
In the proposed experiment, a resonant optical cavity amplifies the intensity fluctuations, and eventually the momentum transferred to the mirror by the bouncing photons. Such fluctuations are actually measured by observing the momentum of the mirror, in particular around a mechanical resonance where its susceptibility increases. The measurement of the mirror motion can be performed interferometrically by a meter field \cite{ref6,ref7,ref8}. 

\begin{figure}[h]
\centering
\includegraphics[width=0.98\textwidth]{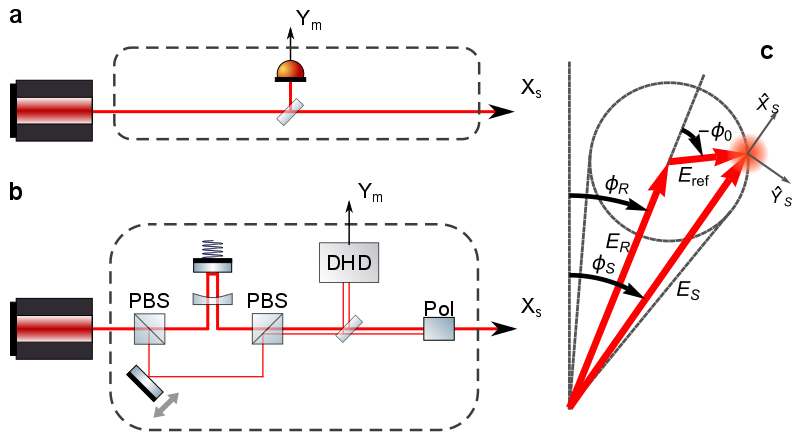}
\caption{Simplified experimental schemes. (a) Scheme of a classical measurement of the field amplitude fluctuations. (b) Simplified experimental setup for our QND measurement. DHD: double homodyne detection; PBS: polarizing beam-splitter; Pol: polarizer. (c) Schematic composition of the fields after the polarizer, in the complex phase plane. The mean field $E_S$ is formed by superposition of the field reflected by the cavity and then transmitted through the polarizer ($E_R$), and a fraction of the reference field ($E_{\mathrm{ref}}$). By changing the length of the reference path we can decide the reference phase $\phi_0$ and actually the output field phase $\phi_s$. The final amplitude quadrature  $X_{\mathrm{s}}$ corresponds to the quadrature  $X^{(\phi_s-\phi_R)}$  in the reflected field.}
\label{fig1}
\end{figure}

The complete measurement apparatus can be viewed as a system with two outputs: the signal field (i.e., a quantum object), and the result of a continuous measurement on one of its quadratures, yielding a (classical) meter variable $Y_{\mathrm{m}}$. An ideal QND measurement is testified by a perfect correlation between the quantum observable to be estimated, i.e. a field quadrature $X_{\mathrm{s}}$ (signal variable), and $Y_{\mathrm{m}}$. The condition to be satisfied can be written as $C_{X_{\mathrm{s}}Y_{\mathrm{m}}} := |S_{X_{\mathrm{s}}Y_{\mathrm{m}}}|^2/(S_{X_{\mathrm{s}}X_{\mathrm{s}}} S_{Y_{\mathrm{m}}Y_{\mathrm{m}}}) = 1$ where $S_{XY}$ is the cross-correlation spectrum between $X$ and $Y$, and $C_{XY}$ is the so-called magnitude-squared coherence (MSC). It is elucidating to compare the QND procedure with a standard, classical intensity measurement where the signal is the quadrature $X_{\mathrm{s}}$ of the field at one output port of a beam-splitter, while the field at the other output port is detected to provide the meter $Y_{\mathrm{m}}$ (Fig. \ref{fig1}a). With a coherent input the cross-correlation is null, and the measurement can just provide information on possible excess noise: the photon noise of the remaining, usable light remains inaccessible. On the contrary, a QND measurement gives access to the quantum fluctuations of the signal field. 

For a deeper understanding of the optomechanical QND measurement, we can consider an ideal scheme exploiting a cavity with coupling rate $\kappa$ and no extra losses, and a resonant input field in a coherent state, whose amplitude ($\Xin$) and phase ($\Yin$) quadrature fluctuations have spectral densities $S_{\Xin \Xin} = S_{\Yin \Yin} = 1/4$.

The position $q$ of the mechanical oscillator embedded in the cavity as end mirror, normalized to its zero-point fluctuations, is given in the Fourier space by
$q = q_{\mathrm{th}} + q_{\mathrm{rp}}$, 
where
$q_{\mathrm{rp}}=4 \, \chi \, \chio \sqrt{\Gba} \, \Xin$
is the displacement due to the radiation pressure, and 
$\Sth = 4 \Gth \,|\chi|^2$
is the displacement spectrum due to the oscillator thermal and quantum noise. 

In the above expressions, 
$\chi = \omega_m/\left(\omega_m^2-\omega^2 - \rmi\omega\gamma_m\right)$
is the mechanical susceptibility,
$\chio = 1/\left(1-\rmi \frac{\omega}{\kappa}\right)$ is the optical susceptibility,
$\Gba = G^2/\kappa$ is the back-action rate \cite{note1}, 
and
$\Gth =  (\omega/Q) (n_T+1/2)$ is the thermal and quantum coupling rate, where $Q$ is the mechanical quality factor \cite{note2} and the average thermal occupancy is 
$n_T = \left(\exp{\left(\frac{\hbar \omega}{kT}\right)}-1\right)^{-1}$.

The field quadratures at the output of the cavity are \cite{ref10}
\begin{eqnarray}
\label{Xo}
\Xo & = & \exp (2 \rmi \phio) \, \Xin    \\
\label{Yo1}
\Yo & = & \exp (2 \rmi \phio) \,\Yin + \sqrt{\Gba} \, \chio \, q \\
&= & \exp (2 \rmi \phio) \,\Yin + 4 \Gba \, |\chio|^2 \, \chi \, \Xo + \sqrt{\Gba} \, \chio \, q_{\mathrm{th}}
\label{Yo}
\end{eqnarray}
where $\phio = \arg[\chio]$. The relations (\ref{Yo1}-\ref{Yo}) describe the interaction between the field to be measured and the optomechanical system. In order to complete a QND measurement of the field, we need an additional readout channel, measuring the oscillator displacement with the result
\begin{equation}
Y_{\mathrm{m}} = q + \, \qr
\label{Eq:qm}
\end{equation}
where $q$ is defined above and $\qr$ is an additional noise term that includes both the readout imprecision and its back-action (i.e., it comprises the overall measurement accuracy). 
In case of detection at the standard quantum limit, the spectrum of $\qr$ is $S^{SQL}_{\qr \qr} = 2 |\chi|$ \cite{Caves}, but the fundamental quantum limit is even lower, i.e., $S^{QL}_{\qr \qr} = 2 \mathrm{Im}[\chi]$ \cite{Jaekel,Kampel}. At the oscillator resonance frequency, the two limits coincide.

From the above model, we extract three meaningful considerations.
(i) Eq. (\ref{Xo}) shows that the optomechanical interaction does not perturb the amplitude field quadrature $\delta X$, that is transmitted to the output.
(ii) Eq. (\ref{Eq:qm}) and the definitions of $q$ and $q_{\mathrm{rp}}$ show that the output of the readout contains some information on $\Xo$.
(iii) Eq. (\ref{Yo}) shows that, in the output field, amplitude and phase quadratures are correlated. 

The first two properties form the basis of the Quantum Non Demolition measurement: $Y_{\mathrm{m}}$ is the result of the QND process, that includes the optomechanical interaction (that does not destroy the variable $\delta X$), and a measurement of $q$. The third observation implies instead that the optomechanical interaction is also producing a field in a squeezed state: since $\Xo$ and $\Yo$ are correlated, there is an output field quadrature $\delta X^{\phi}$ for which the fluctuations are below $S_{\Xo \Xo}$, and eventually below the vacuum level.

Once acquired, $Y_{\mathrm{m}}$ can be used to predict the behavior of the quadrature $\Xo \equiv X_{\mathrm{s}}$ of the surviving field, that is estimated as $\XE = \alpha(\omega) Y_{\mathrm{m}}$, where $\alpha(\omega)$ is an arbitrary complex function that is chosen with the aim of minimizing the average residual uncertainty 
$S_{\Delta X}^{\alpha} := \langle |X_{\mathrm{s}} - \XE|^2 \rangle$.
In a stationary system, the optimal $\alpha$ is $\alpha_{\mathrm{opt}} = (S_{X_{\mathrm{s}} Y_{\mathrm{m}}})^*/S_{Y_{\mathrm{m}} Y_{\mathrm{m}}}$, and the lowest residual uncertainty on the signal is 
$S_{\Delta X} := S_{X_{\mathrm{s}} X_{\mathrm{s}}} \left(1- C_{X_{\mathrm{s}} Y_{\mathrm{m}}} \right)$.

\begin{figure}[h]
\centering
\includegraphics[width=0.7\textwidth]{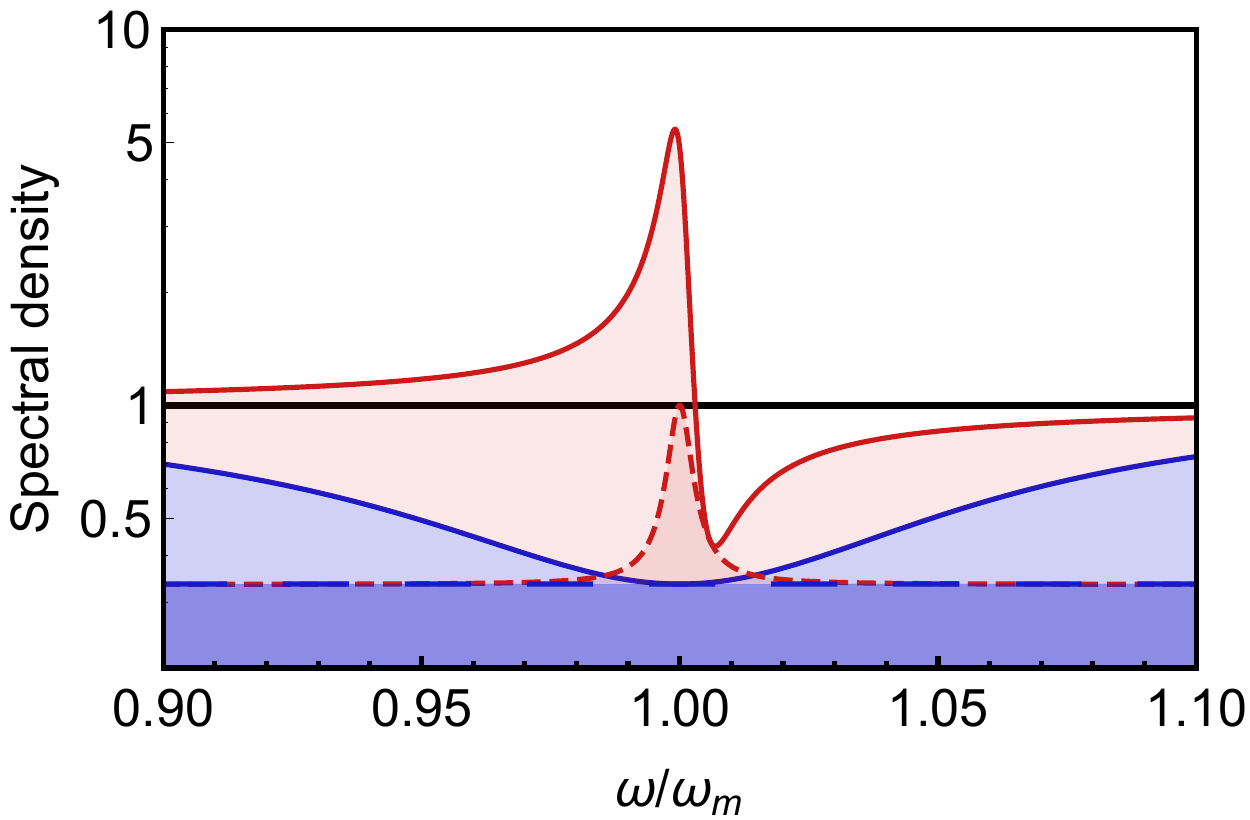}
\caption{Field fluctuations after an optomechanical setup. Blue, long-dashed curve: residual uncertainty $S^{QL}_{\Delta X}$ in the amplitude quadrature of the field, after its QND measurement where the readout of the oscillator motion is performed at the quantum limit. For comparison, with the red dashed line we show the spectral density of an output quadrature $\delta X^\phi$ when the phase $\phi$ is optimized at each frequency. The results of more realistic experiments, with fixed detection phase and readout imprecision, are instead shown with solid lines. In this case, we take $\phi = 0.002$ rad, and we consider a readout at the quantum limit for $\omega=\omega_m$. Spectral densities are normalized to the shot noise level. For all the curves, we consider an input radiation resonant with lossless cavity, in the ``bad cavity'' regime (i.e., with $\chio \simeq 1$). The other optomechanical parameters are inspired by the experiment described below, namely $\gamma_m = 0.005 \,\omega_m$, $\Gba = \,\omega_m$, $\Gth = 0.5 \,\omega_m$.  }
\label{newTheo}
\end{figure}

For the considered optomechanical system, such residual uncertainty can be written as $S^{QL}_{\Delta X} = \left( 1+\frac{\Coop}{1+R} \right)^{-1}$ where $\Coop = \frac{\Gba |\chio|^2}{\Gth}$ is the cooperativity. The parameter $R=\frac{\gamma_m Q}{2 \omega_m (n_T+1/2)}$ is originated by the readout noise $\qr$, considered at the quantum limit, and in general $R \ll 1$ except when the mechanical oscillator is cooled close to its ground state \cite{note3}. An example of this residual spectral density is shown in Fig. \ref{newTheo} with a blue dashed line.

A readout imprecision at the quantum limit requires a rapidly varying detection phase, optimized as a function of the frequency, i.e., a so-called variational readout \cite{ref24}. It is more realistic to consider a QND procedure having a constant, frequency-independent readout imprecision. We can assume that the quantum limit is achieved at the mechanical resonance frequency, and thus set the readout imprecision at $|\chi(\omega_m)| = 1/\gamma_m$. With this choice, we can write the total readout noise as $S_{\qr \qr} = \left(1/\gamma_m + \gamma_m|\chi|^2\right)$, where the second term within brackets is originated by the readout back-action, and $R$ must be multiplied by $\frac{1}{2 \mathrm{Im}[\chi]}\left(\frac{1}{\gamma_m}+\gamma_m |\chi|^2\right) \simeq 1+2\left(\frac{\omega-\omega_m}{\gamma_m}\right)^2$. The resulting residual uncertainty is shown in Fig. \ref{newTheo} with a blue solid line.

It is interesting to compare $S^{QL}_{\Delta X}$ with the spectral density in the maximally squeezed output quadrature $S^{\mathrm{min}}$, that is calculated by minimizing the spectral density of the output field quadrature $\delta X^{\phi}$ with respect to $\phi$, for each detection frequency: $S^{\mathrm{min}} = \frac{1+\sin^2(\arg[\chi])\,\Coop}{1+\Coop}$. Similarly, the residual uncertainty obtained in a QND measurement at fixed readout imprecision can be compared with the noise in a fixed output quadrature $\delta X^\phi$. The two spectra are shown in  Fig. \ref{newTheo} with red lines. A significantly better performance is obtainable with the QND approach, in particular for the realistic experiments using a fixed measurement phase (solid lines). We will further discuss it after the description of our experimental results.

Recent advances in cavity optomechanics \cite{ref10} have allowed to discern the quantum component in the effect of radiation pressure \cite{ref11,ref12,ref13}, and in the correlation between signal and meter \cite{ref12} or between field quadratures after optomechanical interaction \cite{ref14a,ref14,ref14b}. As discussed, the latter is actually the basic ingredient of the observed ponderomotive squeezing \cite{ref14a,ref15,ref16,ref17,ref17a}. The ability of a mechanical oscillator to perform a QND measurement of the radiation intensity fluctuations is experimentally demonstrated with a squeezed microwave source by Clark {\it et al.} \cite{ref17b}, who exploit the phase quadrature of the same driving field, detected after the optomechanical interaction, as a meter of the mirror motion. As discussed, a complete QND scheme requires an additional readout of the mirror motion, without destroying (or perturbing too much) the field to be measured, that is therefore preserved after having disclosed its quantum properties. This is in fact what we are showing in the following, where we describe an experiment that actually achieves a measurement of the transmitted light quantum noise by observing the effect of the photons impact on a movable mirror.

\section{The experiment}
\label{experiment}

In a fair QND procedure, $\alpha (\omega)$ is chosen {\it a priori}, e.g. on the basis of a model, or derived from the analysis of an independent set of data (this analysis could include a destructive measurement of $\Xo$ to estimate its correlation with $Y_{\mathrm{m}}$). In order to verify that a quantum measurement is indeed performed, the experimentalist has to measure the output intensity fluctuations as well as their correlation with $Y_{\mathrm{m}}$. In a realistic optomechanical system, the achievement of an ideal QND is prevented by thermal fluctuations of the movable mirror, by optical losses and by the imprecision of the readout. Moreover, detuning between input field frequency and cavity resonance, and/or excess classical input noise, can create a strong classical correlation between the signal field quadrature $X_{\mathrm{s}}$, and the meter field $Y_{\mathrm{m}}$ \cite{ref8}. Therefore, a non-null correlation $C_{X_{\mathrm{s}}Y_{\mathrm{m}}}$, as it occurs in a classical measurement of a noisy field, is not sufficient to guarantee that an even non-ideal, yet quantum QND measurement is achieved. The model-independent condition to be verified is that the information carried by $Y_{\mathrm{m}}$ is sufficient to reduce the residual uncertainty of $X_{\mathrm{s}}$  below the standard quantum fluctuations (shot noise), i.e., that $S_{\Delta X} < 1$ \cite{ref9,Roch}. 

\begin{figure}[h]
\centering
\includegraphics[width=0.8\textwidth]{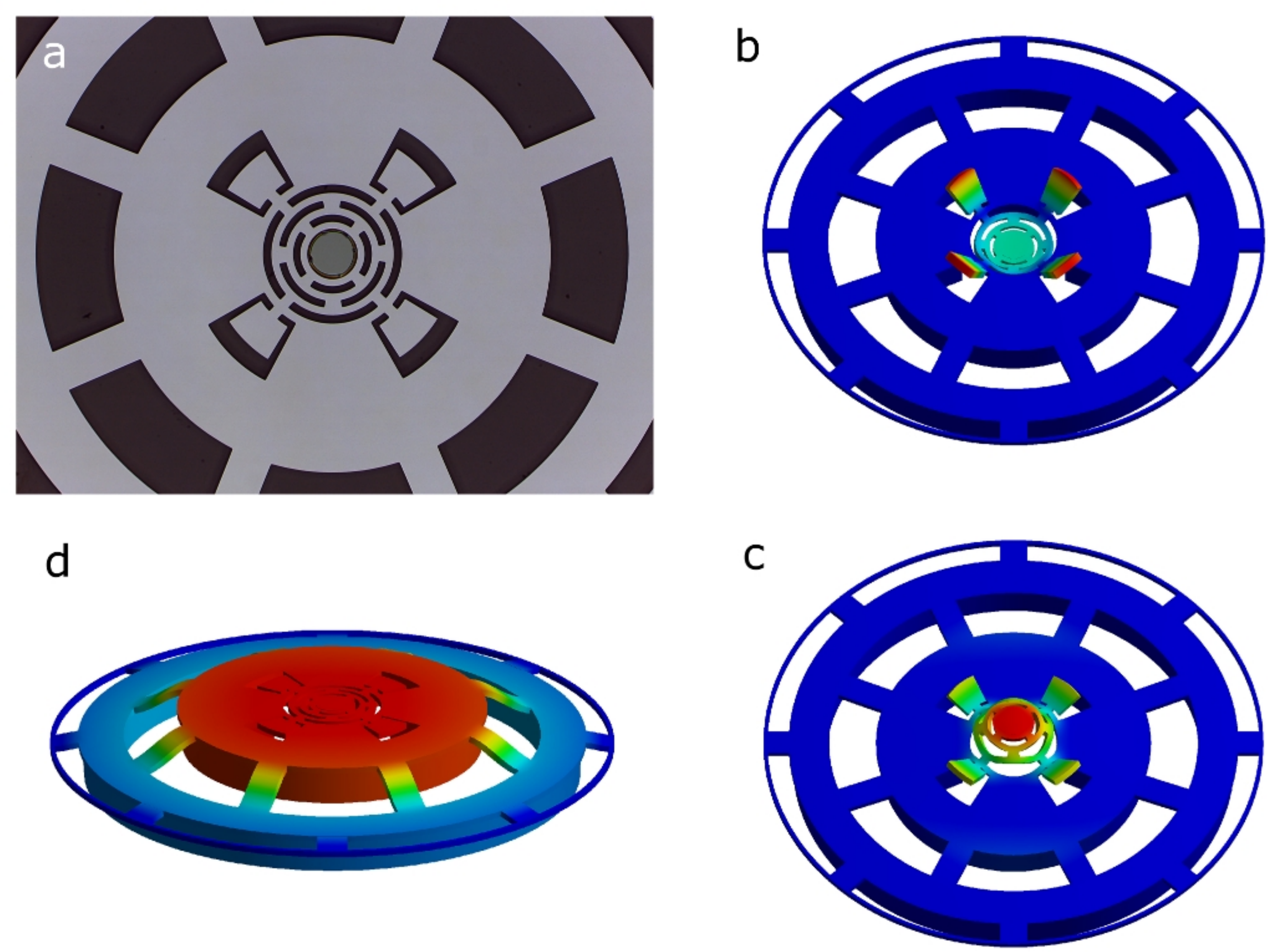}
\caption{The optomechanical oscillator. (a) SEM image of the full device, including the central oscillator and the external isolating wheel. The central dark disk is the $400 \mu$m diameter highly reflective coating. (b-d) FEM simulations of the displacement corresponding to the balanced oscillator mode exploited in this work (b), the second, unbalanced mode (c), and the first wheel oscillator mode (d). 
}
\label{fig5}
\end{figure}

Our experiment is based on an oscillating micro-mirror, working as end mirror in a high finesse optical cavity.
This oscillator is fabricated by micro-lithography on a silicon-on-insulator wafer. A detailed description of the fabrication process is reported in Ref. \cite{ref25}, while the design of the device is discussed in Refs. \cite{ref18,ref19}. The oscillator has a particular shape, studied to maximize its mechanical quality factor and isolation from the frame (Fig. \ref{fig5}a). A structure made of alternating torsional and flexural springs supports the central mirror and allows its vertical displacement with minimal internal deformations, reducing the mechanical loss in the highly dissipative optical coating. For the oscillation mode exploited for this work, the movement of the central disk is balanced by four counterweights, so that the four joints are nodal points (Fig. \ref{fig5}b). Its effective mass is $m = 2.5 \times 10^{-7}$ kg, deduced form the thermal peak in the displacement spectrum measured at room temperature, its frequency at cryogenic temperature is $\omega_{m}/2\pi = 169334$ Hz. The quality factor of $1.1 \times 10^{6}$ at cryogenic temperature is measured in a ring-down experiment. In a second oscillation mode, with resonance frequency around $\sim208$ kHz, the counterweights move in phase with the central disk (Fig. \ref{fig5}c), therefore a net recoil force is applied on the joints, inducing a larger coupling with the frame and actually a lower quality factor. The design includes an external double wheel, working as mechanical filtering oscillator (Fig. \ref{fig5}d), with a resonance frequency of $\sim 22$ kHz.
The central coated region of the oscillator is the back mirror of a $L_{\mathrm{c}} = 1.455$ mm long Fabry-P\'erot cavity where the input coupler is a 50 mm radius concave mirror, glued on a piezoelectric transducer used to keep a cavity resonance within the laser tuning range. A cavity half-linewith of $\kappa/2\pi = 2.85$ MHz is measured at cryogenic temperature. From the calculated Finesse (18055), the measured resonance depth in the reflected intensity, and the measured mode matching of $90\%$, we deduce an input coupler transmission of 315 ppm (in agreement with the direct measurement, input rate $\kappa_1/2\pi = 2.58$ MHz) and additional cavity losses of 33 ppm (loss rate $\kappa_2/2\pi = 0.27$ MHz). The cavity is strongly overcoupled to optimize the quantum efficiency.  
The cavity is suspended inside an helium flux cryostat and thermalized to its cold finger with soft copper foils. The temperature reached by the cavity mount, measured with a diode sensor, is 4.9 K. A finite elements simulation of the heat propagation inside the mount and the silicon device, at the maximum input laser power, suggests that the oscillator temperature should be few tenths of degree higher. The temperature that gives the best agreement between the experimental spectra and the model is indeed 5.6 K. 

The experimental setup is sketched in the simplified scheme of Figure \ref{fig1}b and in more details in Fig. \ref{exp_completo} of Appendix A.
A laser beam from a Nd:YAG source is actively amplitude stabilized, down to an amplitude noise (normalized to shot noise) of 1 + $P$/(24 mW), where $P$ is the laser power. An additional, frequency-shifted  auxiliary beam (not shown in Figure \ref{fig1}) is used for controlling the detuning from the optical cavity (see Appendix A for details). 
The main beam is split by a polarizing beam-splitter (PBS), the outputs of which are sent into the two arms of an interferometer. On one arm, the beam is mode-matched to the optical cavity. The laser power impinging on the cavity is about 50 $\mu$W from the auxiliary beam, and 38 mW from the main beam. The calculated intracavity power is 350 W, corresponding to $n_{\mathrm{c}} = 1.8 \times 10^{10}$ photons. Optical circulators deviate the reflected beams toward the respective detections.

\begin{figure}[h]
\centering
\includegraphics[width=0.9\textwidth]{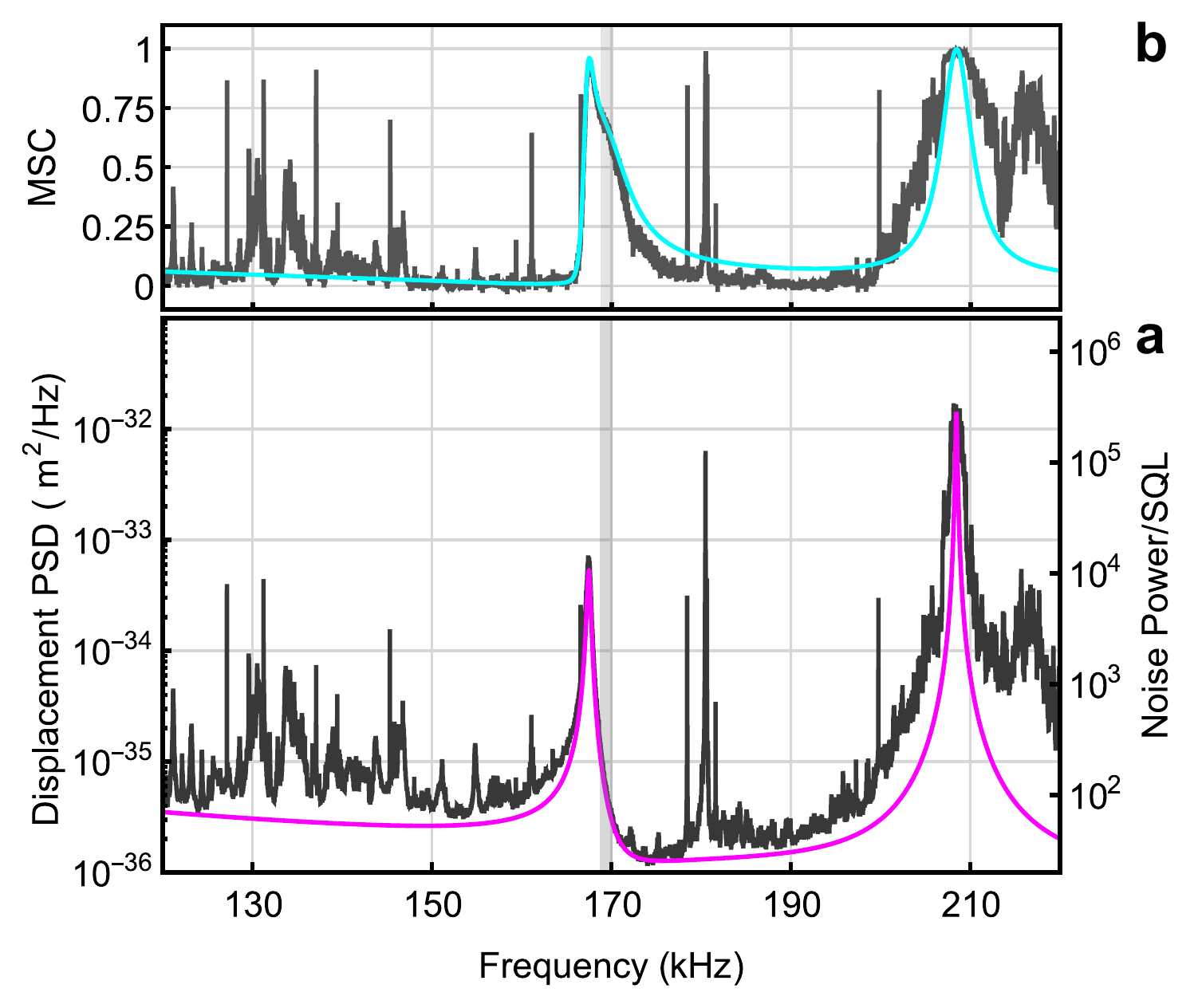}
\caption{Meter spectrum and correlations. (a)  Spectral density of the meter field (black). The spectrum is calibrated both in terms of meter shot noise (SQL; right axis), and in terms of single-sided power spectral density (PSD) of cavity displacement noise (left axis). The electronic noise (already subtracted from the displayed spectrum) is 10 dB below the SQL. In the model (magenta) we have introduced phenomenologically a $1/\omega^2$ contribution to account for the tails of low frequency modes, and an additional resonance at  $\sim 208$ kHz.  (b)  Experimental magnitude-squared coherence $C_{X_{\mathrm{s}}Y_{\mathrm{m}}}$ between the output signal and the meter (black), together with its theoretical model (cyan). The corresponding signal spectrum is shown in Fig. \ref{fig3}. A shadow shows the frequency region where the QND measurement is accomplished.}
\label{fig2}
\end{figure}

After the recombination of the two interferometer beams, a beam sampler picks up about $3\%$ of the p-polarized light arriving from the cavity, and $\sim 15\%$ of the s-polarized light from the reference arm of the interferometer. The collected radiation is detected in two homodyne setups whose output signals, opportunely combined, allow to actively stabilize the interferometer with the desired phase difference between the arms and, at the same time, to derive a weak measurement of the cavity phase noise and actually of the motion of the oscillating mirror (see Appendix A). The meter variable $Y_{\mathrm{m}}$ is obtained in this way without the necessity of an additional readout field, at the expenses of a slightly reduced efficiency in the transmission of the signal beam. The vacuum noise entering from the unused port of the beam sampler determines the measurement imprecision of the readout.
The spectrum of the meter  (Fig. \ref{fig2}a) is dominated by the fluctuations of the cavity length, mainly due to the oscillating mirror. Therefore, the meter provides indeed a readout for the movable mirror, that in turn performs a measurement of a particular field quadrature (namely, the quadrature that gives rise to the intracavity intensity fluctuations).

Due to the weak detuning, the optomechanical interaction shifts the frequency of the main oscillator mode to $\sim 167500$ Hz and broadens its resonance to 430 Hz (corresponding to an effective temperature of  $\sim 2$ mK) \cite{ref10}. With the achieved effective susceptibility, a readout achieving the quantum limited sensitivity at resonance implies an imprecision of $1.5\times 10^{-37}$ m$^2$/Hz
. However, the imprecision level is not visible: the mechanical peak emerges from a background of few $10^{-36}$ m$^2$/Hz given by the tails of low frequency mechanical modes and of the unbalanced oscillator mode at $\sim 208$ kHz. Thermal noise, laser amplitude noise (of classical and quantum origin), and intracavity  intensity fluctuations due to the background displacement noise, all contribute with comparable importance to the force noise acting on the oscillator (see the simulations reported in the Appendix C for their quantitative estimations). The last contribution (i.e., eventually, the cavity phase noise) is responsible for the deviation from a Lorentzian shape of the peak, that assumes a Fano profile. 

The intracavity amplitude quadrature fluctuations are imprinted on the mirror motion. Since the radiation is slightly detuned on the red side of the optical resonance for improving the systems stability, the reflected field quadratures are rotated with respect to the intracavity field, therefore the fluctuations sensed by the oscillator do not exactly correspond to the amplitude quadrature of the reflected field. In order to explore a range of reflected quadratures, we add to the reflected field a small portion of a beam from the reference arm of the interferometer, with a controlled phase (optical path length). 
At this purpose, 
after the beam sampler the main beam is filtered by an high extinction ratio ($>10^7$) polarizer. The axis of the transmitted polarization is very close to the p-polarization axis (within $\sim 1^{\circ}$), so that $> 99\%$ of the field from the cavity and $\sim 3\%$ of the field from the reference arm (corresponding to about 2 $\mu$W) are transmitted and superimposed to form the signal field. The latter is thus rotated with respect to the radiation reflected by the cavity, as outlined in Fig. \ref{fig1}c, with a tuning range of about $\pm 10^{-2}$ rad.

The radiation transmitted by the polarizer is actually the observed physical system, and in particular its amplitude fluctuations are the signal variable $X_{\mathrm{s}}$. In order to verify that the meter provides a QND measurement of such fluctuations, they are monitored (destructively) with a standard balanced detection, composed of a $50\%$  beam-splitter and a couple of photodiodes: the sum of their signals gives $X_{\mathrm{s}}$, their difference provides an accurate calibration of the radiation standard quantum level (SQL). With respect to a standard homodyne detection, this scheme improves the phase stability and, above all, the accuracy of the shot noise calibration, that is not trivial in a homodyne with high signal power, at the price of weak additional losses. 

The measured common mode rejection of the balanced detection is $\sim 40$ dB, and the total quantum efficiency in the detection of the field reflected by the cavity is $69\%$, including the losses in the beam sampler and in the polarizer, and the $\sim90\%$ efficiency of the homodyne photodetectors.

\begin{figure}[h]
\centering
\includegraphics[width=0.95\textwidth]{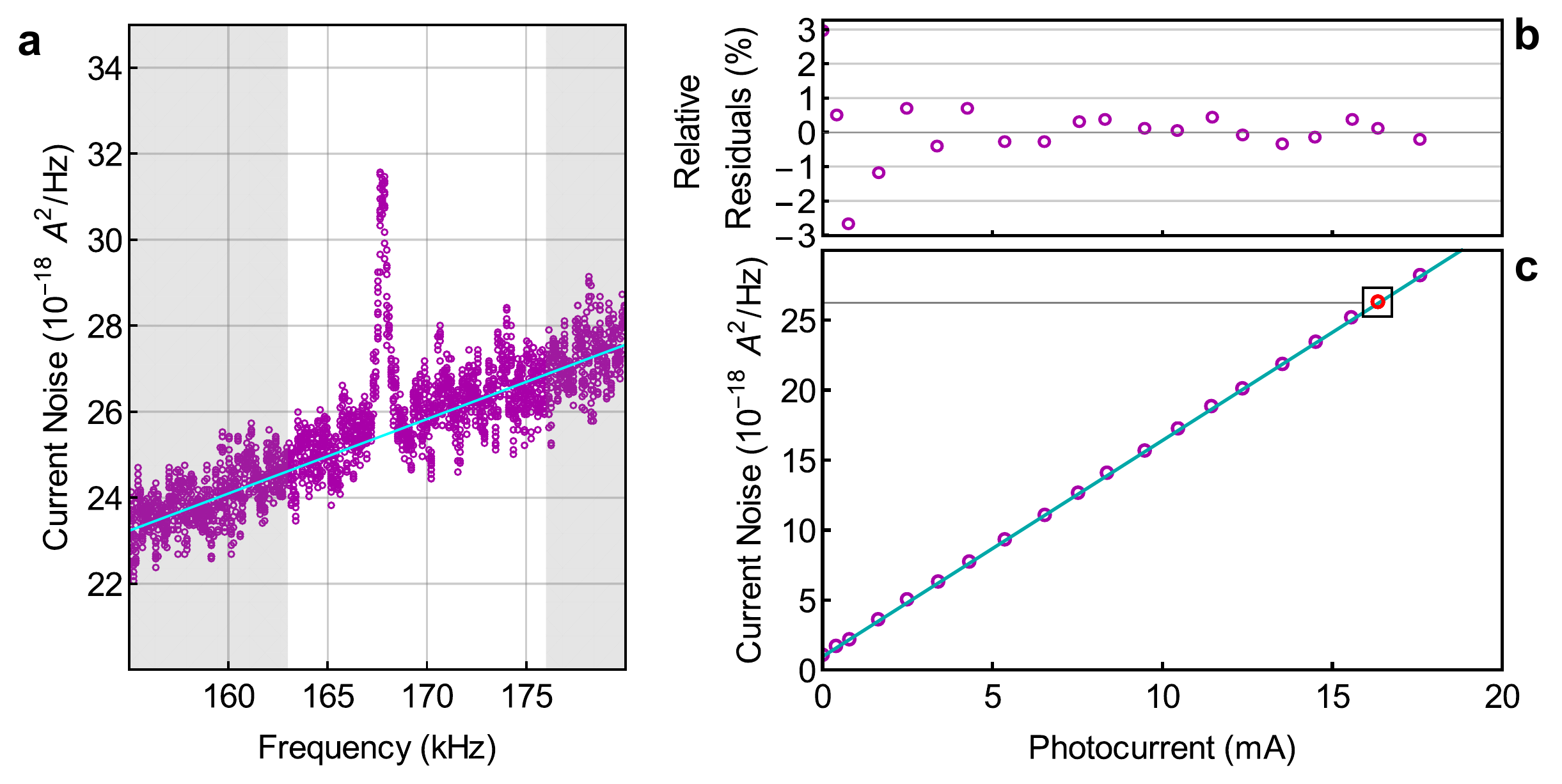}
\caption{Shot noise calibration. (c) Current noise spectral density at 170 kHz measured in the difference signal of the balanced detection, versus total photocurrent, measured by varying the optical power impinging on the detectors with the laser far from resonance. The cyan straight line is a linear fit to the data. The red circle indicates a typical measurement taken with the fully working experiment (with the laser locked to the cavity), used to calibrate the SQL for the spectra reported in Figures \ref{fig3} and \ref{fig4}. (b) Residuals of the fitting procedure. (a) Spectral density of the difference signal, acquired during the experiment. The shadowed region is used for the linear regression shown with a cyan straight line, that is actually used to evaluate the SQL. The peak at $\sim 168$ kHz is the remnant of the signal due to the oscillator.}
\label{fig8}
\end{figure}

The sum and difference signals are filtered with high order low-pass, anti-aliasing circuits and acquired by an high resolution digital scope. The complete electronics for the readout of the sum and difference signals are calibrated with a relative accuracy of better than $0.1 \%$. 
The linearity of the difference signal versus the detected photocurrent (sum of the two detectors photocurrents) has been checked by sending to the photodiodes the laser radiation of the main beam very far from cavity resonance (Figure \ref{fig8}(b-c)). The residuals of the linear fit show no systematic deviation. The noise variance reported in the figure is calculated by considering the spectrum of the photodiodes difference signal in the intervals 154 -- 163 kHz and 176 -- 180 kHz, and calculating the spectral density at 170 kHz with a linear interpolation. Such linear interpolation is sufficient to account for the fact that the spectrum is not flat, due to the filters in the photodetectors circuits. The same procedure is used for evaluating the SQL in the experimental data, where we exclude in this way the region (163 -- 176 kHz) where the strong oscillator peak could percolate into the difference signal in spite of the high rejection. The electronic noise is equal to the shot noise of 0.63 mA, equivalent to an impinging power of 0.8 mW, and it has a day-to-day reproducibility of $\sim 10\%$. Since it is subtracted from the measured spectra, it contributes to the uncertainty with an additional $0.3 \%$. Taking into account all the analyzed sources of systematic error, we evaluate that their total effect in the calibration of the spectra to the SQL is below $0.5 \%$.

\section{Results and discussion}
\label{discussion} 
 
To verify the claim that the amplitude quadrature of the signal field is measured in a QND way by the mechanical oscillator \cite{ref17b}, and actually that the meter variable well reports the result of this measurement, we have to calculate the residual spectrum $S_{\Delta X}$ and show that it falls below the standard quantum level in a proper frequency range. We observe indeed that the coherence between the meter and the signal (Fig. \ref{fig2}b) reaches values close to unit around the peak frequency, but it mainly reflects classical fluctuations. Only the following comparison with the signal spectrum allows to assess that a QND measurement is indeed performed.

\begin{figure}[h]
\centering
\includegraphics[width=0.7\textwidth]{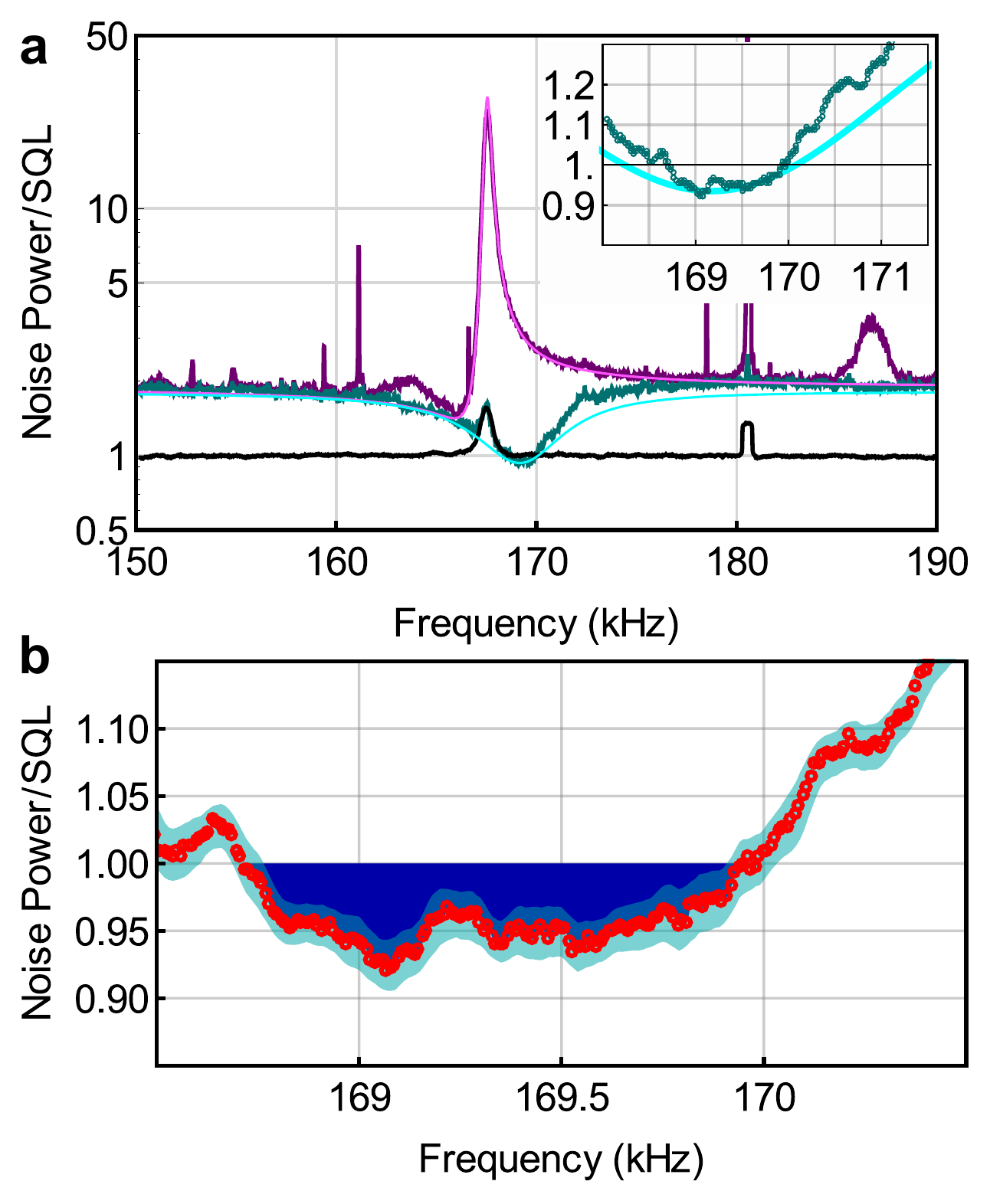}
\caption{Signal and its residual uncertainty. (a)  Spectral density of the signal field $S_{X_{\mathrm{s}}X_{\mathrm{s}}}$, normalized to the SQL (wine). From the comparison with the model (magenta) we deduce a detuning of  $-0.016 \kappa$ and a signal phase $\phi_s = -24$ mrad. Spectral density of the residual fluctuations  $S_{\Delta X}$  (dark green), with their model (cyan).  Spectrum of the difference between the signals of the photodiodes in the balanced detection, from which the SQL is deduced (black). The electronic noise (already subtracted from the displayed spectra) is 15 dB below the SQL, for both the sum and the difference signals of the balanced detection. The inset shows an enlarged view.  (b)  For  $S_{\Delta X}$  we show the result of a flat moving average over a frequency interval  of 150 Hz  (average value with red symbols, and $90 \%$ confidence belt in light blue). The minimum is $0.921\pm0.012$ (uncertainty corresponding to one standard error). By averaging over a 600 Hz band we obtain $0.942\pm0.006$.}
\label{fig3}
\end{figure}
In Fig. \ref{fig3} we show the spectrum of $X_{\mathrm{s}}$, i.e., of the amplitude fluctuations of the output field that is determined by a particular choice of cavity detuning and interferometer reference phase. It displays a typical Fano profile, due to the interference between amplitude fluctuations of the intracavity field, which act on the mirror via radiation pressure, and the field fluctuations induced by the consequent mirror motion. For the chosen reference phase, such interference is constructive on the right of the resonance, and destructive on the left, due to the change in the sign of the real part of the mechanical susceptibility \cite{ref20}. As a consequence, depending on the frequency, the spectral density can be higher or lower than the input amplitude power spectrum, but we always find it above the SQL do to the excess input amplitude noise. This behavior is indeed predicted by the theory (outlined in the Appendix C), and typically occurs for most of the values of the reference phase. 

If, on the other hand, we exploit the information carried by the meter and calculate the spectral density  $S_{\Delta X}$   of the residual fluctuations, we verify that it falls below the shot noise level in a $\sim 1.5$ kHz broad region, on the high frequency side of the resonance. Its lowest value, normalized to the SQL, is $0.921 \pm 0.012$ (uncertainty corresponding to one standard error) when the spectrum is integrated over 150 Hz (Fig. \ref{fig3}b). By averaging over a 600 Hz band we obtain $0.942 \pm 0.006$, demonstrating a QND measurement with strong statistical significance. The systematic error due to calibration uncertainties is $\pm 0.005$. To fully exploit the information carried by the meter, we have not just used the correlation between  $X_{\mathrm{s}}$ and $Y_{\mathrm{m}}$, but also the one between  $X_{\mathrm{s}}$ and the square of $Y_{\mathrm{m}}$ (see Appendix B). 

We have fitted to the spectrum a complete optomechanical model (described in the Appendix C) where all the system parameters are independently measured, except for the amplitude of the background displacement noise, the detuning and the reference phase. The fact that $S_{\Delta X}$  is below the SQL on just the right hand side of the peak, predicted by the complete model, but in contrast with our simplified introductory description (see Fig. \ref{newTheo}), is due to the favorable frequency/background noise cancellation occurring around the resonance frequency of the free oscillator \cite{ref21}, an effect that can also be interpreted in terms of Opto-Mechanically-Induced Transparency \cite{omit}.  

\begin{figure}[h]
\centering
\includegraphics[width=0.7\textwidth]{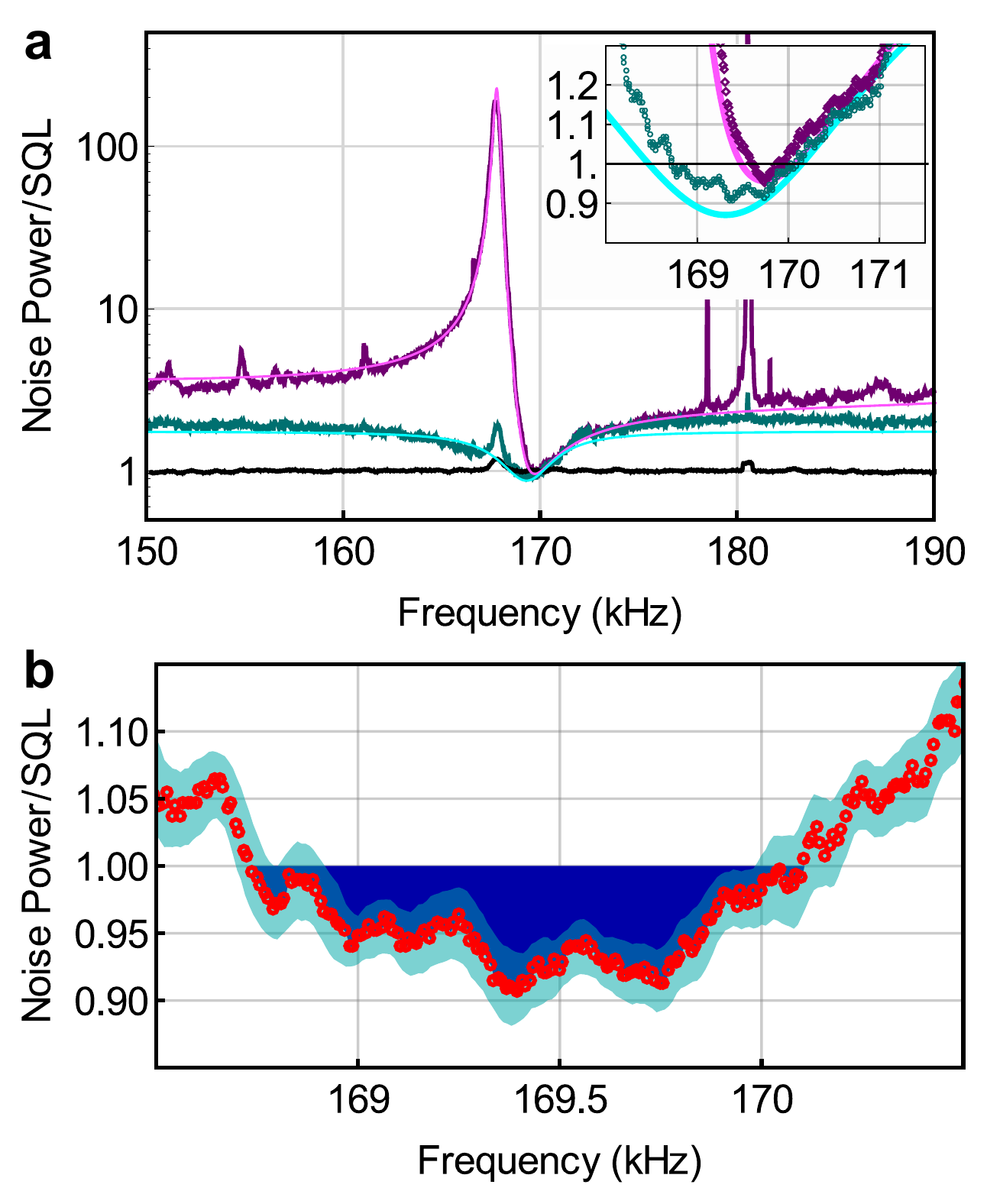}
\caption{Ponderomotive squeezing and QND measurement. With respect to Fig. \ref{fig3}, the signal spectral density and its residual uncertainty are recorded for a different value of the signal phase (-41.5 mrad)  and a slightly different detuning  ($-0.019 \kappa$). The signal field now shows ponderomotive amplitude squeezing, however the information carried by the meter is still useful to enhance the sub-shot noise property of its residual uncertainty. The minimal value of the residual fluctuations is now $0.906\pm0.016$ when integrated over 150 Hz, and $0.924\pm0.007$ integrating over 600 Hz.}
\label{fig4}
\end{figure}
For a deeper exploration of the QND measurement, it is interesting to vary the choice of the signal quadrature $X_{\mathrm{s}}$. Such variation has strong effects on the intensity spectrum $S_{X_{\mathrm{s}}X_{\mathrm{s}}}$. In particular, if the phase $\phi_s$ is changed to the opposite side with respect to the reference given by the reflected field (see Fig. \ref{fig1}c), the destructive interference in $X_{\mathrm{s}}$  occurs on the right of the resonance. By accurately tuning the phase, we can now observe an intensity spectrum $S_{X_{\mathrm{s}}X_{\mathrm{s}}}$  falling below the shot noise level (Fig. \ref{fig4}). It is the signature of ponderomotive squeezing \cite{ref15,ref16,ref17,ref22,ref23}. On the other hand, the residual spectrum obtained after exploiting the information carried by $Y_{\mathrm{m}}$ is weakly phase-dependent, as indicated by the fact that the sub-SQL region is now very similar to the one previously shown in Fig. \ref{fig3}. 

Both the dependence of  $S^{\mathrm{min}}$ from $\arg[\chi]$ and the limited squeezing bandwidth in a given output quadrature are the consequence of the physical origin of the squeezing, that is due to the negative interference (cancellation) between the terms $\Xo  \cos \phi$ and $\propto  \chi  \, \Xo  \sin \phi$ (see Eq. (\ref{Yo})) in the output field quadrature $\delta X^{\phi}$. At a given phase $\phi$, such interference is optimal for a particular value of $\chi$, i.e., for a particular frequency, while it degrades as soon as $\chi$ varies. Moreover, the cancellation is limited by the imaginary part of the second term (actually, by the imaginary part of $\chi$), and it is completely inefficient at resonance, where $\chi$ is purely imaginary (see the dashed red curve in Fig. \ref{newTheo}). Such limiting features are absent in the QND measurement, where an appropriate weighting function $\alpha(\omega)$ can compensate for the frequency dependence of $\chi$ and for its argument. On the other hand, as we have seen, the QND needs an additional measurement (on $q$) besides the optomechanical interaction, that is not necessary to produce the squeezed field. The QND performance depends on the quality of such measurement. 

\section{Conclusions}

In conclusion, we have experimentally demonstrated that a QND measurement is performed by means of the mechanical interaction of light with a moving mirror. More specifically, our optomechanical apparatus produces a radiation field whose amplitude fluctuations (including those of quantum origin) are continuously observed. The result of such measurement is available through a meter channel that actually monitors the mirror motion. The back-action of the measurement is almost completely confined to the signal field phase fluctuations, and its weak percolation in the amplitude quadrature is efficiently detected by the meter. As a consequence, the residual fluctuations of the signal amplitude, that remain unknown after exploiting the information brought about by the meter, are below the shot noise. In a measurement process, the SQL is a crucial threshold: one can reduce the noise down to the SQL by just using, in a noise eater, a beam sampler to measure the intensity fluctuations. On the contrary, in a classical apparatus a noise level below the shot can just be obtained inside the close loop containing the detector, i.e., in a destructive measurement. In other words, the quantum photon noise remains elusive in classical experiments, and it can just be catch by a QND measurement \cite{ref9}. This technique is therefore very promising for the application to sub-SQL sensors, including integrated micro-devices and future gravitational wave detectors, where it can either be used to produce sub-shot noise light in a quantum noise eater \cite{Yamamoto}, or directly integrated in the complete measurement procedure by performing a preliminary estimation of the field quantum fluctuations.

This work has been supported by MIUR ("PRIN 2010-2011" and "QUANTOM") and by INFN ("HUMOR" project). A.B. acknowledges support from the MIUR under the "FIRB Futuro in ricerca 2013" funding program, project code RBFR13QUVI.

\section*{Appendix A: The Pound-Drever-Hall and the double homodyne detections.}

\begin{figure}[h]
\centering
\includegraphics[width=0.99\textwidth]{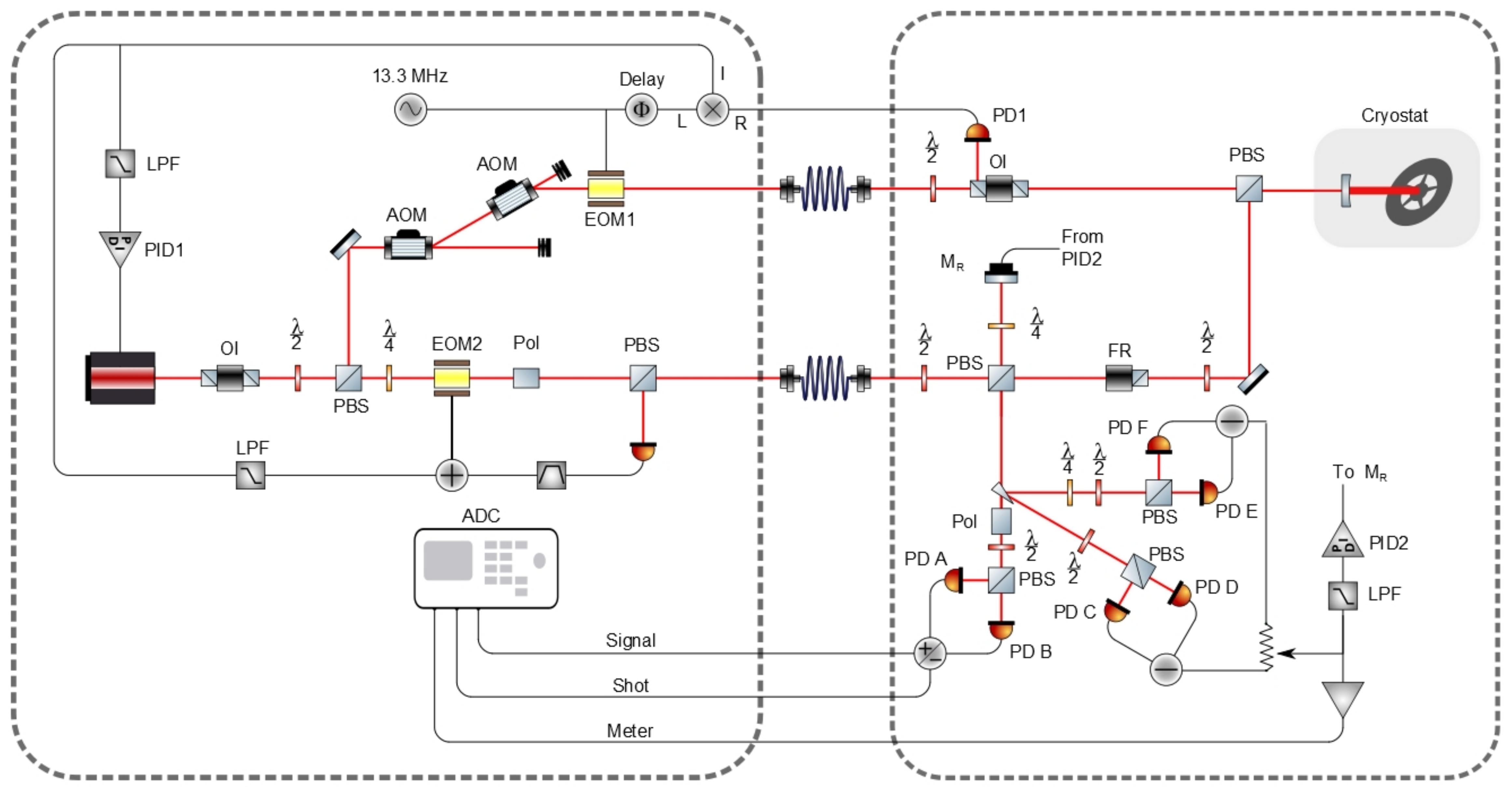}
\caption{Detailed scheme of the experimental setup. EOM: electro-optic modulator. AOM: acousto-optic modulator. Pol: polarizer. PBS: polarizing beam-splitter. LPF: low-pass filter. OI: optic isolator. PD: photodiode. FR: Faraday rotator.}
\label{exp_completo}
\end{figure}

The experimental setup is shown in Fig. \ref{exp_completo}.
On the laser bench, the laser radiation is split into two beams. The first one (auxiliary beam) is frequency shifted by means of two acousto-optic modulators (AOM) operating on opposite diffraction orders. A resonant electro-optic modulator (EOM) provides phase modulation at 13.3~MHz used for a Pound-Drever-Hall (PDH) detection scheme. The PDH signal allows to stabilize the laser frequency to the cavity resonance. The locking bandwidth is about 20 kHz and additional notch filters assure that the servo loop does not influence the system dynamics in the frequency region around the oscillator frequency. 

The PDH signal is also bandpass filtered around 22 kHz, and added to the signal driving the intensity modulator of the noise eater acting on the main beam. We so implement a feedback cooling \cite{ref26} on the wheel oscillator, with two purposes: firstly, we improve its dynamic stability, that is otherwise critical due to the combined effect of optomechanical interaction and frequency locking servo loop \cite{ref27}. Secondly, we depress the fluctuations of the wheel oscillator, that would otherwise provide a major contribution to the overall cavity phase noise. We remind that the rms value of such phase noise is large enough that a simple linear expansion of the cavity optical response in not sufficient to account for the reflected field fluctuations. Therefore, even if feedback cooling is just effective on the peak of the wheel resonator, it reduces the contribution brought into the frequency range of interest by nonlinear mixing.    

The second beam (main beam) is actively amplitude stabilized, reducing the noise by about 30 dB in the band 100kHz - 200 kHz. Both beams are sent to the experiment bench by means of single-mode, polarization maintaining optical fibers. The main beam is split by a polarizing beam-splitter (PBS), the outputs of which are sent into the two arms of a Michelson interferometer. The length of the reference arm is finely controlled by shifting its end mirror with an inductive transducer. On the other arm, the beam is overlapped to the auxiliary beam, with orthogonal polarizations, in a further PBS and then mode-matched to the optical cavity. 
On the path of the radiation exiting from the Michelson interferometer, the two faces of a wedge window, with the bisector plane at the Brewster angle, pick up $1.5\%$ each of the p-polarized light arriving from the cavity, and respectively $6\%$ and $23\%$ of the s-polarized light from the reference beam. On the path of one of these reflections, a quarter-wave plate with the axes parallel to the polarizations adds an additional delay between the fields arriving from the cavity and the reference arms. The fields reflected by the two window faces are analyzed by homodyne setups, each composed of a half-wave plate that rotates the polarizations by $45^{\circ}$, a PBS, and a couple of photodiodes at the two outputs of the PBS. The difference signals of the two couples of photodiodes can be written respectively as $\,V_A \sin \phi\,$ and $\,V_B \cos \phi$, where $\phi$ is the phase difference between the fields coming from the two arms of the Michelson interferometer. The transimpedence gains of the detectors are set to compensate for the different collected powers, in order to have $V_A \simeq V_B$. We electronically derive a weighted average of the two signals $\,V_\mathrm{m} = \beta_1 V_A +\beta_2 (1-\beta_1) V_B \propto \sin(\phi + \phi_0(\beta_1,\beta_2))$, where $\beta_1$ can be chosen between 0 and 1, and $\beta_2 = \pm 1$, so that $-\pi/2 < \phi_0 < \pi/2$. The low frequency component of $V_\mathrm{m}$ is integrated and fed back to the position control of the reference arm mirror, so that the phase $\phi$ is locked to $-\phi_0$ with a servo bandwidth of about 1 kHz. Moreover, the fluctuations of $\delta V_\mathrm{m}$ are now proportional to the fluctuations of the phase quadrature of the field reflected by the cavity, plus a contribution that, due to the low detected power, can be considered as originated by additional vacuum fluctuations. In summary, such combined homodyne detection is equivalent to a standard homodyne detection of the phase quadrature of the radiation reflected from the cavity, and it additionally allows to choose and stabilize the phase difference between the two, orthogonally polarized fields that compose the transmitted main beam. We identify $\delta V_\mathrm{m}$ with our meter variable $Y_\mathrm{m}$. 

\section*{Appendix B: Data acquisition and analysis}

We have acquired simultaneous data streams from three channels: the sum and difference outputs from the final balanced detection, and the meter $Y_{\mathrm{m}}$. The signals are sampled at 5 MHz and several 10 seconds data streams are acquired, separated by lapses of few seconds necessary for data storage. Such delays improve the randomness of the complete data sets, reducing the effect of long term relaxations. The stability of the mean beam power is better than $1 \%$ during the whole measurement period. The data elaborated to obtain the results shown in Figures \ref{fig3} and \ref{fig4} are taken respectively from 5 and 4 consecutive 10 s time series.  

The 10 seconds temporal series are divided into 100 ms long intervals. A preliminary selection on the intervals is performed by setting upper limits on the peak and rms values of the sum signal. This selection procedure is useful to reject datasets plagued by strong noise spikes, mainly due to low ($\sim$kHz) frequency modes, generated by instabilities of the helium flux in the cryostat. We keep $\sim90 \%$ of the data intervals. 

For each $n$-th interval, we calculate the discrete Fourier transform of the difference signal $\tilde{X}_{-}^{(n)}$, of the sum signal $\tilde{X}_{+}^{(n)}$, of the meter signal $\tilde{Y}_{\mathrm{m}}^{(n)}$, of the square of the meter signal $\tilde{Y}_{\mathrm{sqm}}^{(n)}$ (we distinguish in the following the experimental signal $\tilde{X}_{+}$ from the signal variable $X_\mathrm{s}$ that is obtained from $\tilde{X}_{+}$ after subtraction of the electronic noise and normalization to the SQL).
 
The spectra to be evaluated are the sum and difference power spectra $S_{X_+X_+}$ and $S_{X_-X_-}$, and the cross-correlation contributions. For all such spectra, we use correct estimators as discussed below in the sub-section ``The statistical estimators". 
The final steps of the analysis are the subtraction of the detection electronic noise, and the normalization of the sum and the residual spectra to the SQL. The obtained $S_{X_\mathrm{s}X_\mathrm{s}}$ is plotted in Figures \ref{fig3} and \ref{fig4} (wine traces).

\subsubsection*{Correlation with the square of the meter}

Due to the relatively large rms value of the cavity phase noise, mainly due to several mechanical resonances, a simple linear expansion of the cavity reflection function is not sufficient to account for the whole effect of such fluctuations on the reflected field quadratures. As a consequence, the best estimate of $X_{\mathrm{s}}$ would be a function $f(Y_{\mathrm{m}})$. If we consider its second order expansion, we deduce that a non-null correlation can also exist between $X_{+}$ and the square of $Y_{\mathrm{m}}$, and a more accurate estimate of the signal state can be performed by exploiting all the information provided by the meter signal, i.e., using an appropriate linear combination of $Y_{\mathrm{m}}$ and of its square. This residual uncertainty is found by subtracting from $S_{X_+X_+}$ also the correlation between $X_{+}$ and $Y_{\mathrm{sqm}}$. This is indeed the spectrum of the residual fluctuations that we have plotted in Figures \ref{fig3} and \ref{fig4} (dark green traces). It is compared with the theoretical calculation of $S_{\Delta X}$, that is based on linear expansions of the equations of motion. We remark however that even without the use of the correlation with $Y_{\mathrm{sqm}}$, the normalized spectrum of the residual fluctuations falls below the unit. No further improvements have been obtained by considering correlation with higher order in $Y_{\mathrm{m}}$.  

\begin{figure}[h]
\centering
\includegraphics[width=0.98\textwidth]{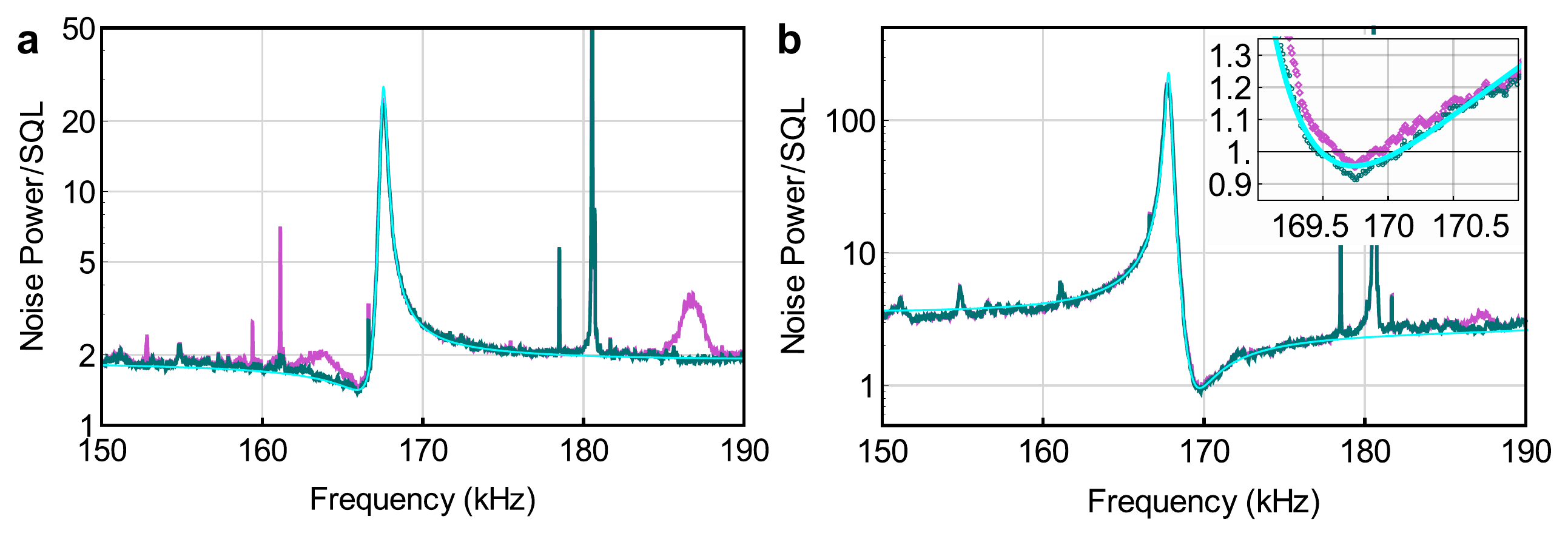}
\caption{(a) Spectra of the signal (experimental spectrum of the sum signal from the homodyne detection, with the electronic noise subtracted, and normalized to SQL) (purple), the same after subtraction of the correlation with the square of the meter (dark green), and theoretical model (cyan), for the reference phase corresponding to Figure \ref{fig3}. (b) The same, for the reference phase corresponding to Figure \ref{fig4}. The inset displays an enlarged view around the minimum, showing the improvement of the ponderomotive squeezing when the correlation with $Y_{\mathrm{sqm}}$ is subtracted from the spectrum.}
\label{fig9}
\end{figure}

Even the subtraction of just the correlation with $Y_{\mathrm{sqm}}$ from the spectrum $S_{X_+X_+}$ is interesting, as shown in Fig. \ref{fig9}, since it removes some peaks originating from the nonlinearity of the system, improving the agreement with the model. We point out that this is a confirmation of the existence of a quadratic nonlinearity. The correlation with $Y_{\mathrm{sqm}}$ is particularly meaningful around two peaks at $\sim163$ kHz and $\sim187$ kHz, but it also slightly improves the residual around the minimum. 

\subsubsection*{The statistical estimators}

We have to estimate power spectra (such as $S_{X_{+}X_{+}}$ and $S_{X_{-}X_{-}}$), as well as cross-correlation contributions (such as $|S_{X_+Y_\mathrm{m}}|^2/S_{Y_\mathrm{m}Y_\mathrm{m}}$), starting from a finite number $N$ of experimental, Fourier transformed time series. For the power spectrum of a variable $X$, a good estimator is straightforwardly $\hat{S}_{XX} = \sum_{n=1,N} |\tilde{X}^{(n)}|^2/N\,$. On the other hand, finding a correct, unbiased indicator for the cross-correlation contribution is not obvious. We have therefore chosen a different point of view.

We are willing to estimate the residual fluctuations of $X$ that remains once the information brought by $Y$ is optimally used (the subscripts of $X$ and $Y$ are omitted in this discussion for the sake of clarity). In a linear system, the information that can be extracted from $Y$ can be written as $\alpha(\omega) \tilde{Y}$, where $\alpha(\omega)$ is a complex function. Therefore, we have to find the function $\alpha(\omega)$ that minimizes the spectral density of $S_{\Delta X}^{\alpha} := \langle |\tilde{X} - \alpha \tilde{Y}|^2 \rangle = S_{XX} + |\alpha|^2 S_{YY} - 2 \mathrm{Re} (\alpha S_{XY})$. By deriving with respect to $\alpha$, we find that its optimal value is 
$\alpha_{\mathrm{opt}} = (S_{XY})^*/S_{YY}$ 
and the lowest residual spectrum is indeed 
$S_{\Delta X}^{\mathrm{opt}} = S_{XX}-|S_{XY}|^2/S_{YY}$. 
Any different $\alpha$ gives an overestimation of the optimal residual spectrum. On the other hand, for a given $\alpha$, we have a correct, unbiased estimator of the residual spectrum, that is 
\begin{equation}
\hat{S}_{\Delta X}^{\alpha} = 1/N \left(\sum |\tilde{X}^{(n)}|^2 + |\alpha|^2 \sum|\tilde{Y}^{(n)}|^2 - 2 \mathrm{Re}\left(\alpha \sum (\tilde{X}^{(n)})^* \tilde{Y}^{(n)}\right) \right)  \,. 
\label{stima}
\end{equation}
The function $\alpha$ could be chosen {\it a priori}, e.g. on the basis of a model, but for a more realistic analysis we have derived it from the experimental data using the definition of $\alpha_{\mathrm{opt}}$ as guideline, as described in the following. We separate the $N$ intervals into two independent half-sets, according to the parity of the index $n$. From the first half-set we calculate $\alpha$ as 
$\alpha_{\mathrm{odd}} = \sum_{\mathrm{odd}\, n} \tilde{X}^{(n)} (\tilde{Y}^{(n)})^*/\sum_{\mathrm{odd}\, n}|\tilde{Y}^{(n)}|^2$, 
and from the second half-set we calculate the residual spectrum following Eq. (\ref{stima}), where the sums are taken over the even indexes. We then repeat the procedure by exchanging the two half-sets, and we finally take the average over the two resulting residual spectra. If we calculate the expectation value of our final spectrum $S_{\mathrm{\Delta X}}^{\mathrm{exp}}$, we find:
\begin{displaymath}
\begin{array}{lcl}
E\left[ S_{\Delta X}^{\mathrm{exp}} \right]  & 
= & \frac{1}{N} \langle \, \sum_{\mathrm{even}\, n} |\tilde{X}^{(n)}|^2 + |\alpha_{\mathrm{odd}}|^2 \sum_{\mathrm{even}\, n}|\tilde{Y}^{(n)}|^2 - 2 \mathrm{Re}(\alpha_{\mathrm{odd}} \sum_{\mathrm{even}\, n} (\tilde{X}^{(n)})^* \tilde{Y}^{(n)}) \\
& & \quad +  \, (\mathrm{even} \leftrightarrow \mathrm{odd}) \, \rangle  
\\
& = & S_{XX} + \langle |\alpha_{\mathrm{e/o}}|^2 \rangle  S_{YY} - 2 \mathrm{Re} \left( \langle \alpha_{\mathrm{e/o}} \rangle S_{XY} \right)  
\end{array}
\end{displaymath}
where we have used the independence of the two half-data sets, so that, e.g.,  $\,\langle \alpha_{\mathrm{odd}} \sum_{\mathrm{even}\, n} f^{(n)} \rangle = \langle \alpha_{\mathrm{odd}} \rangle \langle \sum_{\mathrm{even}\, n} f^{(n)} \rangle \, $ and $\, \langle \alpha_{\mathrm{odd}} \rangle = \langle \alpha_{\mathrm{even}} \rangle := \langle \alpha_{\mathrm{e/o}} \rangle$. Since $ \, \langle |\alpha_{\mathrm{e/o}}|^2 \rangle \ge |\langle \alpha_{\mathrm{e/o}} \rangle|^2 \,$ (a relation valid for any stochastic variable), we can write
$E\left[ S_{\Delta X}^{\mathrm{exp}} \right] \,\ge\,  S_{\Delta X}^{\langle\alpha_{\mathrm{e/o}}\rangle}  \,\ge\, S_{\Delta X}^{\mathrm{opt}}$.
Therefore, our experimental evaluation of the residual spectrum provides an unbiased, conservative estimator of the residual spectrum. 

\begin{figure}[h]
\centering
\includegraphics[width=0.9\textwidth]{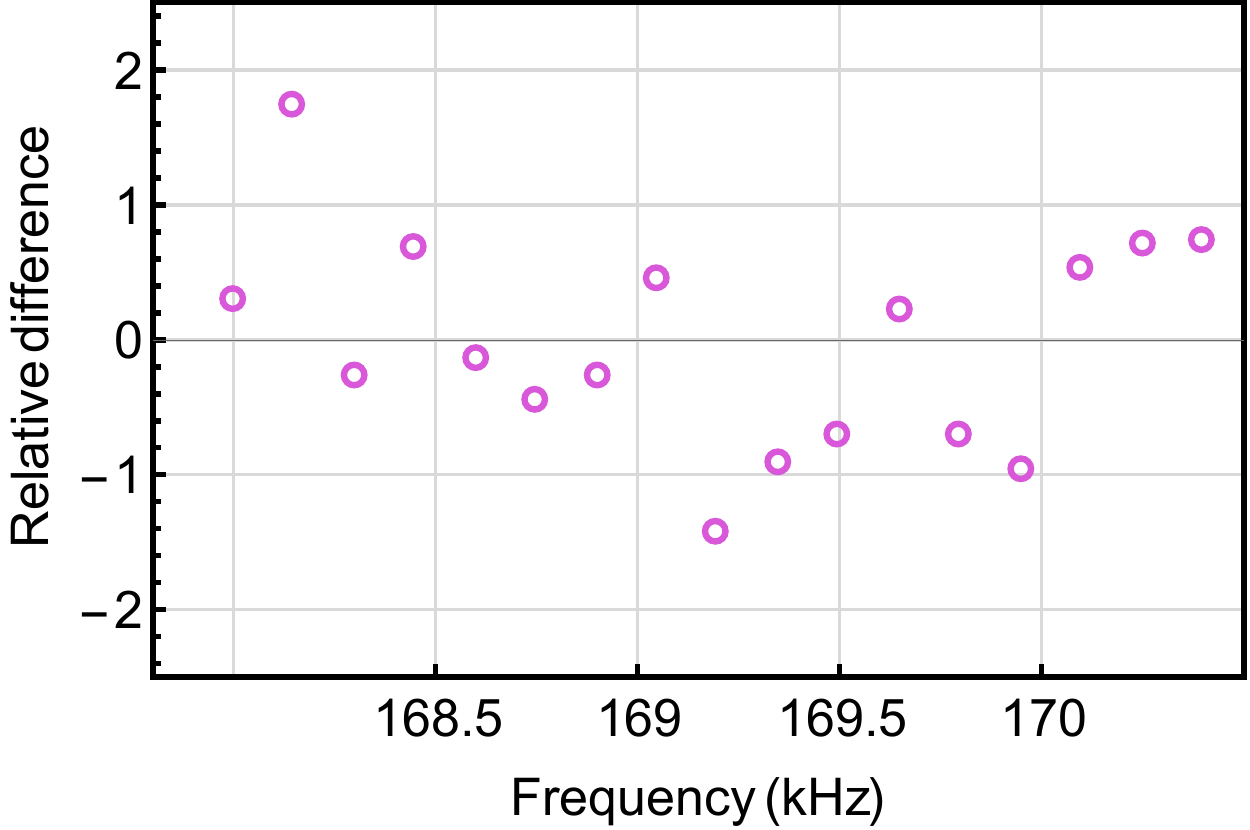}
\caption{Difference between the ``odd" and ``even" estimates of the residual fluctuations, normalized to its statistical uncertainty.}
\label{chk}
\end{figure}
We have tested the compatibility of the two independent ``odd"/``even" estimates by calculating their difference normalized to its statistical uncertainty (i.e., to twice the standard deviation of their average). The result is shown in Fig. \ref{chk} for the QND frequency region of Fig. \ref{fig4}b. The normalized differences have an average value of $-0.024\pm0.21$ and a standard deviation of $0.86\pm0.21$, figures compatible with a normal distribution.
   
The above discussion can be extended to the case of two information channels $Y_1$ and $Y_2$, as follows. We have to find the functions $\alpha_1$ and $\alpha_2$ that minimizes the spectrum of $\,(X - \alpha_1 Y_1 - \alpha_2 Y_2)\,$. The residual spectrum is 
\begin{equation}
S_{\Delta X} = S_{XX}+|\alpha_1|^2 S_{Y_1Y_1}+ |\alpha_2|^2 S_{Y_2Y_2}-2 \mathrm{Re}\left(\alpha_1 S_{XY_1} \right)-2 \mathrm{Re}\left(\alpha_2 S_{XY_2} \right)+2 \mathrm{Re}\left(\alpha_1^{*} \alpha_2 S_{Y_1Y_2} \right)
\label{Sxyz}
\end{equation}
and the optimal weight functions are
\begin{equation}
\alpha_{1,\mathrm{opt}} = \frac{S_{Y_2Y_2}S^*_{XY_1}-S^*_{XY_2} S_{Y_1Y_2}}{S_{Y_1Y_1} S_{Y_2Y_2}-|S_{Y_1Y_2}|^2}
\end{equation}
\begin{equation}
\alpha_{2,\mathrm{opt}} = \frac{S_{Y_1Y_1}S^*_{XY_2}-S^*_{XY_1} S_{Y_2Y_1}}{S_{Y_1Y_1} S_{Y_2Y_2}-|S_{Y_1Y_2}|^2}   \, .
\end{equation}
As in the case of the single correlation, in order to derive a correct estimator one can separate the data streams into two interlaced subsets, calculate the weight functions from the above expression using, in place of the spectra, averages on half-sets of the correspondent discrete Fourier transforms, calculate the residual spectra $S_{\Delta X}$ according to Eq. (\ref{Sxyz}), exchange the two subsets, and finally average the two results.

In our experiment, we have one single meter $Y_{\mathrm{m}}$. However, as we have already mentioned, a linear approximation is not sufficient to fully exploit it. The best estimate of $X_{\mathrm{s}}$ would be a function $f(Y_{\mathrm{m}})$, that we can ideally expand to the second order, as $f(Y_{\mathrm{m}}) \simeq \alpha_1(\omega) Y_{\mathrm{m}} + \alpha_2 (\omega) Y_{\mathrm{sqm}}$, thus returning to the previous, two channels case.
In the ideal case of an infinite number of measurements in a stationary system, the addition of further orders in $Y_{\mathrm{m}}$ can just improve the estimate. However, in the case of $N$ measurements each further channel adds statistical uncertainty. Moreover, the non-optimal estimator can even increase the residual spectrum if the correlation is not sufficiently strong. As already mentioned, in our case we have indeed verified that the residual spectrum is not further improved by considering higher order terms in $Y_{\mathrm{m}}$.

\section*{Appendix C: The model}

The Hamiltonian of the optomechanical system can be written as
\begin{equation}
H=\hbar\omega_{\mathrm{c}}a^{\dagger}a+\frac{1}{2}\hbar\omega_{m}(p^{2}+q^{2})
-\hbar G_{0}a^{\dagger}a q 
\label{hamiltonian}
\end{equation}
where $a$ is the annihilation operator of the cavity mode at frequency $\omega_{\mathrm{c}}$, $p$ and $q$ are the momentum and position operators of the mechanical oscillator, the single-photon coupling strength is $G_0 =-(\omega_{\mathrm{c}}/L_{\mathrm{c}})\sqrt{\hbar/m \omega_m}$. 

The evolution equations for the system are derived from the Hamiltonian with the inclusion of an intense laser field at frequency $\omega_{0}$, input vacuum field operators $a_1^{\mathrm{in}}$  (from the input mirror) and $a_2^{\mathrm{in}}$ (from cavity losses), and additional noise terms that will be listed below. They can be written in the frame rotating at the laser frequency, that is detuned by $\Delta_0 = \omega_0 - \omega_{\mathrm{c}}$ with respect to the cavity resonance, as
\begin{eqnarray}
\dot{q}&=&\omega_{m} p, \label{equazioniq}\\
\dot{p}&=&-\omega_{m} q - \gamma_{m} p + G_0 a^{\dagger}a + \xi, \\
\dot{a}&=&-\kappa a +\rmi \left(\Delta_0+\fn+G_0 q \right)a   +E_0 \nonumber \\
&& +\sqrt{2\kappa_1}\left( a_1^{\mathrm{in}}+\epsilon\right)+\sqrt{2\kappa_2} a_2^{\mathrm{in}} \, .\label{equazionia}
\end{eqnarray}
Here $E_0=\sqrt{2 \kappa_1 P/\hbar \omega_0}$ where $P$ is the input laser power and we take $E_0$ as real, which means that we use the driving laser as phase reference for the optical field.
The mechanical mode is affected by a viscous force with damping rate $\gamma_{m}$ and by a Brownian stochastic force $\xi(t)$. 
We have included the laser excess amplitude noise with the real stochastic variable $\epsilon$. The additional cavity phase fluctuations are introduced by a stochastic term $\fn$ in the detuning. The input fields correlations are
\begin{eqnarray}
&& \bigl\langle a_j^{\mathrm{in}}(t)a_j^{\mathrm{in}}(t')\bigr\rangle = \bigl\langle a_j^{\mathrm{in}, \dagger}(t)a_j^{\mathrm{in}, \dagger}(t')\bigr\rangle = \bigl\langle a_j^{\mathrm{in},\dagger}(t)a_j^{\mathrm{in}}(t')\bigr\rangle= 0,\quad \label{corr1}\\
&& \bigl\langle a_j^{\mathrm{in}}(t)a_j^{\mathrm{in},\dagger}(t')\bigr\rangle = \delta (t-t'), \,\,\, j=1,2.\label{corr2}
\end{eqnarray}

We consider the motion of the system around a steady state characterized by the intracavity electromagnetic field of amplitude $\alpha_s$, and the oscillator at a new position $q_s$, by writing:
\begin{eqnarray}
 q&=&q_s+\delta q,\label{eq:ss+fluct_q}\\
 p&=&p_s+\delta p,\label{eq:ss+fluct_p}\\
 a&=&\alpha_s+\delta a.\label{eq:ss+fluct_a}
\end{eqnarray}

Substituting Eqs.~(\ref{eq:ss+fluct_q})-(\ref{eq:ss+fluct_a}) into Eqs.~(\ref{equazioniq})-(\ref{equazionia}), we obtain the stationary solutions:
\begin{eqnarray}\label{eq:ss_0th}
 q_s&=&\frac{G_0}{\omega_m}|\alpha_s|^{2},\\
 p_s&=&0,\\
 \alpha_s&=&\frac{E_{0}}{\kappa-\rmi\Delta}, \label{eq:ss_1th}
\end{eqnarray}
where $\Delta = \Delta_0 + G_0 q_s$, and the first order linearized  equations for the fluctuations operators 
\begin{eqnarray}
\delta \dot{q}&=&\omega_m \delta p,\label{linearq}\\
 \delta \dot{p}&=&-\omega_m \delta q-\gamma_m \delta p+G_0\left(\alpha_s\delta a^{\dagger}+\alpha_s^{*} \delta a \right)+\xi,\\
 \delta \dot{a}&=&-\left(\kappa-\rmi\Delta\right)\delta a+\rmi G_0\alpha_s\delta q+\sqrt{2\kappa_1}(a_1^{\mathrm{in}}+\epsilon) + \rmi \alpha_s \fn + \sqrt{2\kappa_2}a_2^{\mathrm{in}}.
 \label{lineara}
\end{eqnarray}

The Fourier transformed of Eqs. (\ref{linearq})-(\ref{lineara}), are solved for $a(\omega)$ (we call $a(\omega)$ the Fourier transformed of $\delta a(t)$ and $a^{\dagger}(\omega)$ the Fourier transformed of $\delta a^{\dagger}(t)$, with the same notation for the other fields). 
Using the input/output relations
\begin{eqnarray}\label{eq:output}
E_\mathrm{R}&=&\sqrt{2\kappa_1}\alpha_s-\frac{E_0}{\sqrt{2 \kappa_1}} , \\
a^{\mathrm{out}}&=&\sqrt{2\kappa_1}\delta a-\left(a_1^{\mathrm{in}}+\epsilon\right)  ,
\end{eqnarray}
we can write the output field, with average value
\begin{equation}
E_\mathrm{R}=\sqrt{\frac{P}{\hbar \omega_0}}\frac{\kappa-2\kappa_2+\rmi\Delta}{\kappa-\rmi\Delta}
\label{ER}
\end{equation}
and fluctuation operator
\begin{eqnarray}\label{eq:aout}
  a^{\mathrm{out}}(\omega) &=& \nu_1(\omega) a_1^{\mathrm{in}}(\omega) + \nu_2(\omega)a_1^{\mathrm{in},\dagger}(\omega) + \nu_3(\omega)a_2^{\mathrm{in}}(\omega)+\nu_4(\omega)a_2^{\mathrm{in},\dagger}(\omega)\nonumber \\
  &&+ \nu_5(\omega) \fn(\omega)+\nu_6(\omega)\epsilon (\omega) + \nu_{7}(\omega) \xi (\omega) \, ,
\end{eqnarray}
where
  \begin{eqnarray*}\label{eq:nu1}
  \nu_1(\omega) &=& \frac{\kappa-2\kappa_2+\rmi\bigl(\Delta+\omega\bigr)}{\kappa-\rmi\bigl(\Delta+\omega\bigr)}+\frac{\rmi |G|^2 \kappa_1 \chi_\mathrm{eff}(\omega)}{\bigl[\kappa-\rmi\bigl(\Delta+\omega\bigr)\bigr]^2}, \\
  \label{eq:nu2}
  \nu_2(\omega) &=& \frac{\rmi G^2 \kappa_1 \chi_\mathrm{eff}(\omega)}{\bigl[\kappa-\rmi\bigl(\Delta+\omega\bigr)\bigr]\bigl[\kappa+\rmi\bigl(\Delta-\omega\bigr)\bigr]}, \\
  \nu_3(\omega) &=&\sqrt{\frac{\kappa_2}{\kappa_1}}\left(\nu_1(\omega)+1\right), \\
   \nu_4(\omega) &=&\sqrt{\frac{\kappa_2}{\kappa_1}}\nu_2(\omega), \nonumber\\
   \label{eq:nu5}
   \nu_5(\omega)&=&\frac{\rmi\alpha_s}{\sqrt{2\kappa_1}}\left(\nu_1(\omega)-\nu_2(\omega)+1 \right), \\
   \nu_6(\omega)&=&\nu_1(\omega)+\nu_2(\omega),  \\  
  \nu_{7}(\omega) &=& \frac{i G \sqrt{\kappa_1} \chi_\mathrm{eff}(\omega)}{\kappa-\rmi\bigl(\Delta+\omega\bigr)}\,.
  \end{eqnarray*}
Here $G=G_0 \sqrt{2} \alpha_s$ is the effective coupling strength, and
\begin{equation}
\chi _\mathrm{ eff}(\omega )=\omega _{m}\Biggl[\omega_{m}^{2}-\omega^{2}-\mathrm{i}\omega \gamma _{m}+\frac{|G|^2\Delta\omega _{m}}{\bigl(\kappa -\mathrm{i}\omega \bigr)^{2}+\Delta^{2}}\Biggr]^{-1} \label{chieffD}
\end{equation}
is the effective mechanical susceptibility modified by the optomechanical coupling.

In the experiment, we split the output field into a weak meter and a signal. They have different optical losses, that are considered in the model using the beam-splitter relations
\begin{eqnarray}
a_{\mathrm{m}} = \sqrt{\eta_{\mathrm{m}}} \,a^{\mathrm{out}} +  \sqrt{1-\eta_{\mathrm{m}}} \,a_{\mathrm{3}}  \\
a_{\mathrm{s}} = \sqrt{\eta_{\mathrm{s}}} \,a^{\mathrm{out}} +  \sqrt{1-\eta_{\mathrm{s}}} \,a_{\mathrm{4}} 
\end{eqnarray}
where $a_{3,4}$ are vacuum input fields and $\eta_{\mathrm{m,s}}$ are the efficiencies respectively for the meter and the signal. The correlation between $a_3$ and $a_4$ could be considered by introducing in the model the beam-splitter that separates the meter and signal fields, followed by further beam-splitters modeling the optical losses. However, due to the low efficiency $\eta_{\mathrm{m}}$, to reproduce the results we can safely neglect such correlation and consider vacuum fields $a_{3,4}$ satisfying the relations (\ref{corr1})-(\ref{corr2}) with $j$ extended to (3,4).
We can similarly consider the reference field as contributing to the meter and the signal with independent effective vacuum fields, already included phenomenologically in $a_{3,4}$. To account for the non perfect mode matching we must consider that the field in the non-resonant modes is reflected by the cavity input mirror, and impinges on the detectors where, in first approximation, it does not interfere with the main mode. Therefore, we do not sum the fields, but the fluctuating intensities. The above relations are modified by replacing $\,P \rightarrow \eta_{mm} P, \,$ $\,(a_1^{\mathrm{in}}+\epsilon) \rightarrow \sqrt{\eta_{mm}}\left(a_1^{\mathrm{in}}+\epsilon\right)+\sqrt{1-\eta_{mm}}\,a_5\,$ and $\,a^{\mathrm{out}} \rightarrow \sqrt{\eta_{mm}} a^{\mathrm{out}} + \sqrt{1 - \eta_{mm}} \left(\sqrt{1 - \eta_{mm}} (a_1^{\mathrm{in}}+\epsilon) - \sqrt{\eta_{mm}} a_5\right)$ where $\eta_{mm}$ is the mode matching coefficient and $a_5$ is a further vacuum input field.

The general quadrature of a field $a$ is defined as $\, a \mathrm{e}^{-i \phi}+a^{\dagger}\mathrm{e}^{i \phi}$. For the meter field, we measure the phase quadrature with respect to the field reflected by the cavity. The latter, according to Eq. (\ref{ER}), is dephased by $\,\phi_\mathrm{R} = \arctan \Delta/(\kappa-2\kappa_2) + \arctan \Delta/\kappa \,$ with respect to our reference (i.e., the field at the cavity input). The measured quadrature of the meter is therefore defined by $\,\phi_{\mathrm{m}} = \phi_\mathrm{R} +\pi/2. \,$ Concerning the signal, we are defining as $X_\mathrm{s}$ the amplitude quadrature at the output of the polarizer, i.e., the quadrature defined by the superposition of main and reference fields: $\,\phi_{\mathrm{s}} = \phi_{\mathrm{R}} - \arcsin \left(\sin \phi_0/\sqrt{1+P_\mathrm{R}/P_{\mathrm{ref}}+2\sqrt{P_\mathrm{R}/P_{\mathrm{ref}}} \cos \phi_0}\right) \,$ where $P_\mathrm{R}\,$  ($P_{\mathrm{ref}}$) is the power transmitted by the polarizer and coming from the cavity (reference) arm (Fig. 1c of the main text). 

Theoretical curves are obtained by calculating symmetrized power spectra and cross-correlation spectra of the variables $\,Y_{\mathrm{m}}= a_{\mathrm{m}}(\omega) \mathrm{e}^{-\rmi \phi_{\mathrm{m}}}+a_{\mathrm{m}}^{\dagger}(\omega)\mathrm{e}^{\rmi \phi_{\mathrm{m}}}\,$ and $\,X_{\mathrm{s}} = a_{\mathrm{s}}(\omega) \mathrm{e}^{-\rmi \phi_{\mathrm{s}}}+a_{\mathrm{s}}^{\dagger}(\omega)\mathrm{e}^{\rmi \phi_{\mathrm{s}}}.\,$ The oscillator and cavity parameters, quoted in main text, are all measured independently and fixed in the theoretical calculations. The calculated coupling strengths are $G_0/2\pi = -3.85$ Hz and $G/2\pi = -740$ kHz (at resonance). The input power is $P = 38$ mW. The spectrum $S_{\epsilon\epsilon}$ is 1/4 of the excess intensity noise, normalized to SQL. In our case, we set $S_{\epsilon\epsilon}=0.25\times P/(24 \mathrm{mW})$. The stochastic term in the detuning is linked to the cavity length fluctuations $\delta l$ by $\zeta = \delta l \times \omega_{\mathrm{c}}/L_{\mathrm{c}}$. Its spectrum is modeled with a Lorentzian peak at 208 kHz that roughly reproduces the resonance of the second oscillator mode, plus a $1/\omega^2$ background. The total background amplitude is left as free fitting parameters. The mode-matching parameter and the efficiencies, both for the signal and for the meter, are measured independently. The detuning and the signal phase $\phi_{\mathrm{s}}$ are free fitting parameters. 

\begin{figure}[h]
\centering
\includegraphics[width=0.9\textwidth]{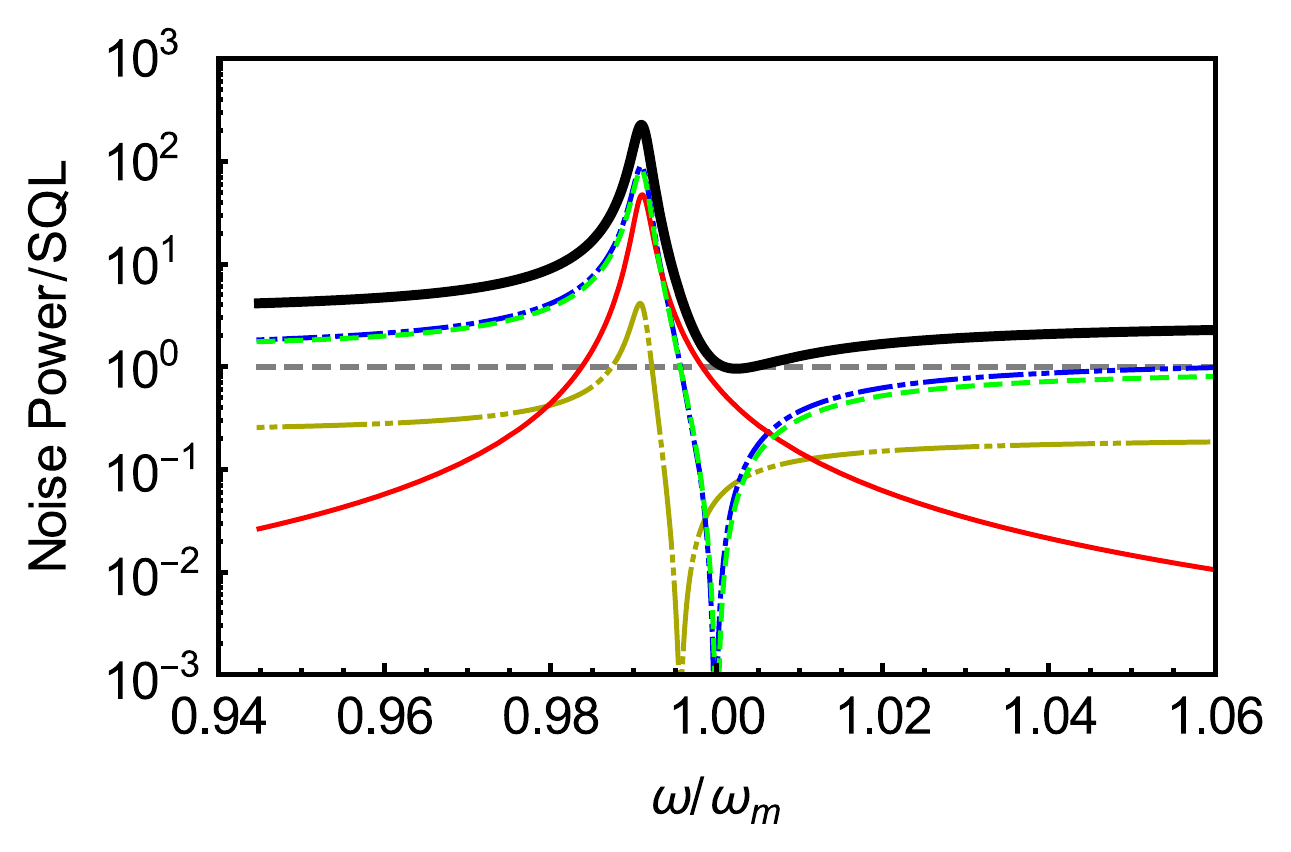}
\caption{Theoretical calculation of the spectrum reported in Figure 7 of the main text (black solid line), normalized to SQL (grey dashed line), together with its contributions: input laser noise (blue dashed-double dotted line), thermal noise (red solid line), vacuum noise entering through cavity losses (dark yellow, long dash-double dotted line), cavity phase noise (greed dashed line).}
\label{fig11}
\end{figure}

In addition to the curves already compared with the experimental results in the main text, we show in Figure \ref{fig11} the contributions of the different noise sources to the power spectrum $S_{X_\mathrm{s}X_\mathrm{s}}$ plotted in Fig. 7 of the main text. 
The contribution of the cavity phase noise cancels at the bare oscillator frequency, as discussed in Ref. [34] of the main text. The contribution of the laser noise (quantum noise and classical amplitude noise) reaches a minimum at a frequency determined by the best destructive interference between the fluctuations of the laser field (modified by the optical cavity) and those mediated by the optomechanical interaction (originated by the term proportional to $\delta q$ in Eq. (\ref{lineara})). With the parameters used for this spectrum, even this interference occurs close to $\omega_m$. This coincidence allows to observe ponderomotive squeezing, that would otherwise be hidden by the cavity phase noise. The squeezing depth is eventually limited by thermal noise.

\begin{figure}[h]
\centering
\includegraphics[width=0.9\textwidth]{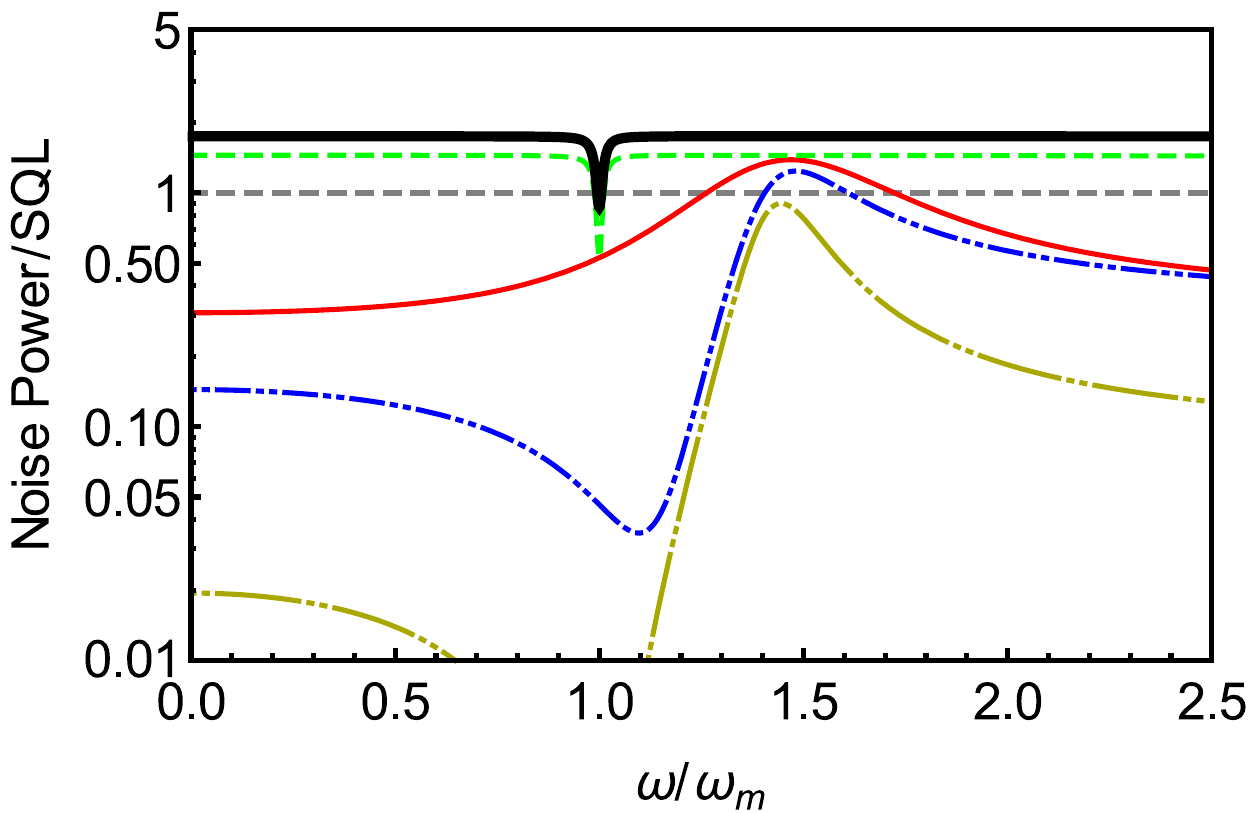}
\caption{Theoretical calculation of the spectrum of the residual fluctuations of $X_\mathrm{s}$, reported in Figure 7 of the main text (black solid line), normalized to SQL (grey dashed line). The other curves show the contributions of the different noise sources to the final spectrum. We start by a system with just the laser noise, at zero temperature, without cavity losses and without extra displacement noise (dark yellow, long dash-double dotted line). We then add cavity losses (blue dashed-double dotted line), thermal noise (red solid line), cavity phase noise (greed dashed line). The inclusion of vacuum noise introduced by the detection efficiency brings to the final spectrum.}
\label{fig12}
\end{figure}
In Figure \ref{fig12} we show, for the same parameters, the spectrum of the residual fluctuations of $X_\mathrm{s}$ after the subtraction of the correlation with the meter. To put into evidence the different noise contributions, we start by the residual spectrum where just the laser noise is present, than we add cavity losses, thermal noise and cavity phase noise. Before the last contribution, the interference effect above described is no more necessary to fall below the SQL, and the region where it happens is potentially much larger. However, in agreement with the comment to the previous figure, we see that eventually the cavity phase noise strongly limits the width of this QND region. Its cancellation at $\omega_m$ is crucial, while the minimum of the spectrum is again limited by thermal noise.


\begin{thebibliography}{99}
\bibitem{note1} The gain $G$ is here defined as $G=2 \sqrt{\frac{\hbar}{2m\omega_m}} \,\frac{\omega_{\mathrm{c}}}{L_{\mathrm{c}}} \sqrt{n_{\mathrm{c}}}$ where $m$ is the oscillator mass, $L_{\mathrm{c}}$ the cavity length, $\omega_{\mathrm{c}}$ the cavity resonant frequency, $n_{\mathrm{c}}$ the cavity photon number at resonance.
\bibitem{note2} We keep a distinction between the oscillator natural linewidth $\omega_m/Q$ and its effective width $\gamma_m$ that can be modified by optical cooling.
\bibitem{note3} We notice that $R$ can be written as $R=1/2 n_{\mathrm{eff}}$, where $n_{\mathrm{eff}}$ is the effective occupation number of the mechanical oscillator.
\end{thebibliography}
\end{document}